\documentclass[aps,nofootinbib,preprintnumbers]{revtex4}
\usepackage{CJK}
\usepackage{lipsum}
\usepackage{amsfonts}
\usepackage{amsmath}
\usepackage{amssymb}
\usepackage[english]{babel}
\usepackage{graphicx}
\usepackage{epsfig}
\usepackage{bm}
\usepackage{verbatim}
\usepackage[utf8]{inputenc}
\usepackage{booktabs}
\usepackage{multirow}
\usepackage{subfig}
\usepackage{slashed}
\usepackage{xcolor}
\usepackage[colorlinks=true,urlcolor=red,citecolor=red]{hyperref}
\usepackage[font=small]{caption}
\usepackage{float}
\usepackage{blindtext}
\usepackage{placeins}
\usepackage[utf8]{inputenc}
\usepackage{bbm}
\DeclareMathOperator{\trace}{tr}

\newcommand{\mathsym}[1]{{}}

\newcommand{\be}{\begin{equation}}
\newcommand{\ee}{\end{equation}}
\newcommand{\bea}{\begin{eqnarray}}
\newcommand{\eea}{\end{eqnarray}}
\newcommand{\ben}{\begin{enumerate}}
\newcommand{\een}{\end{enumerate}}
\newcommand{\bit}{\begin{itemize}}
\newcommand{\eit}{\end{itemize}}
\newcommand{\bde}{\begin{widetext}}
\newcommand{\ede}{\end{widetext}}

\newcommand{\bc}{\begin{center}}
\newcommand{\ec}{\end{center}}

\DeclareUnicodeCharacter{2212}{-}
\input{tcilatex}
\begin{document}

\title{Fermion mass hierarchy and g-2 anomalies in an extended 3HDM Model}
\author{A. E. C\'arcamo Hern\'andez$^{{a,b,c}}$}
\email{antonio.carcamo@usm.cl}
\author{Sergey Kovalenko$^{{b,c,d}}$}
\email{sergey.kovalenko@unab.cl}
\author{M. Maniatis$^{{e}}$}
\email{maniatis8@gmail.com}
\author{Ivan Schmidt$^{{a,b,c}}$}
\email{ivan.schmidt@usm.cl}
\affiliation{\hspace{2cm}\\
$^{{a}}$Universidad T\'ecnica Federico Santa Mar%
\'{\i}a, Casilla 110-V, Valpara\'{\i}so, Chile\\
$^{{b}}$Centro Cient\'{\i}fico-Tecnol\'ogico de Valpara\'{\i}so, Casilla 110-V, Valpara\'{\i}so, Chile\\
$^{{c}}$Millennium Institute for Subatomic Physics at the High-Energy
Frontier, SAPHIR, Calle Fern\'andez Concha No 700, Santiago, Chile\\
$^{{d}}$Departamento de Ciencias F\'isicas, Universidad Andr\'es Bello, 
Sazi\'e 2212, Piso 7, Santiago, Chile\\
$^{{e}}$Departamento de Ciencias B\'asicas, UBB, Casilla 447, Chill\'an,
Chile,\\
}
\date{\today }

\begin{abstract}
We propose an extension of the three-Higgs-doublet model (3HDM), where the
Standard Model (SM) particle content is enlarged by the inclusion of two
inert $SU_{2L}$ scalar doublets, three inert and two active electrically neutral gauge singlet scalars, charged vector like fermions and Majorana neutrinos. These
additional particles are introduced to generate the SM fermion mass
hierarchy from a sequential loop suppression mechanism. In our model the top
and exotic fermion masses appear at tree level, whereas the remaining
fermions get their masses radiatively. Specifically, bottom, charm, tau and
muon masses appear at 1-loop; the masses for the light up, down and strange
quarks as well as for the electron at 2-loop and masses for the light active neutrinos at 3-loop.
Our model successfully accounts for SM fermion masses and mixings and
accommodates the observed Dark Matter relic density, the electron and muon
anomalous magnetic moments, as well the constraints arising from charged
Lepton Flavor Violating (LFV) processes. The proposed model predicts charged LFV decays within the
reach of forthcoming experiments.\newline

\footnotesize
DOI:\href{https://doi.org/10.1007/JHEP10(2021)036}{10.1007/JHEP10(2021)036}
\normalsize
\end{abstract}

\maketitle

\section{Introduction}
\label{themodel}

Despite the great consistency of the Standard Model with experimental data,
it has several unexplained shortcomings. 
Among the most pressings are the absence of any explanation for the
smallness of the masses of the neutrinos and the electron, and for the
existence of three fermion families, accompanied by its mixing. The huge
fermion mass hierarchy, which spreads over a range of 13 orders of
magnitude, from the light neutrino mass scale up to the top quark mass,
lacks any explanation. Moreover, there is no assertion for the smallness of
the quark mixing angles, which contrasts with the sizable values of two of
the three leptonic mixing angles. 

To tackle the limitations of the SM, various extensions, including larger
scalar and/or fermion sectors as well as extended symmetries, discrete and
(or) continuous, with radiative seesaw mechanisms, have been proposed in the
literature \cite{Balakrishna:1988ks, Ma:1988fp, Ma:1989ys, Ma:1990ce,
Ma:1998dn, Kitabayashi:2000nq, Ma:2006km, Dong:2006gx, Chang:2006aa,
Gu:2007ug, Ma:2008cu, Sierra:2008wj, Nardi:2011jp, Huong:2012pg,
Restrepo:2013aga, Ma:2013yga, Ma:2013mga, Hernandez:2013mcf,
Hernandez:2013dta, Okada:2013iba, Sierra:2014rxa, Campos:2014lla,
Boucenna:2014ela, Hernandez:2015hrt, Aranda:2015xoa, Restrepo:2015ura,
Longas:2015sxk, Fraser:2015zed, Fraser:2015mhb, Okada:2015bxa, Wang:2015saa,
Sierra:2016qfa, Arbelaez:2016mhg, Nomura:2016emz, Nomura:2016fzs,
Kownacki:2016hpm, Kownacki:2016pmx, Nomura:2016ezz, Camargo-Molina:2016yqm,
Camargo-Molina:2016bwm, vonderPahlen:2016cbw, Bonilla:2016diq, Gu:2016xno,
Nomura:2017emk, Nomura:2017vzp, Nomura:2017tzj, Wang:2017mcy,
Bernal:2017xat, CarcamoHernandez:2017kra, CarcamoHernandez:2017cwi,
Ma:2017kgb, Cepedello:2017eqf, Dev:2018pjn, CarcamoHernandez:2018hst,
Rojas:2018wym, Nomura:2018vfz, Reig:2018mdk, Bernal:2018aon, Calle:2018ovc,
Aranda:2018lif, Cepedello:2018rfh, CarcamoHernandez:2018vdj, Ma:2018zuj,
Ma:2018bow, Li:2018aov, Arnan:2019uhr, Ma:2019byo, Ma:2019iwj,
CarcamoHernandez:2019xkb, Ma:2019yfo, Nomura:2019yft, Nomura:2019vqc,
Nomura:2019jxj, CentellesChulia:2019gic, Bonilla:2018ynb, Pramanick:2019oxb,
Arbelaez:2019wyz, Avila:2019hhv, CarcamoHernandez:2019cbd,
CarcamoHernandez:2019lhv, Arbelaez:2019ofg, CarcamoHernandez:2020pnh,
CarcamoHernandez:2020ney, CarcamoHernandez:2020pxw,
CarcamoHernandez:2020owa, CarcamoHernandez:2020ehn,
Hernandez:2021uxx,CarcamoHernandez:2021iat}. Furthermore, several theories
with enlarged particle spectrum and symmetries have been constructed to
explain the experimental value of the muon anomalous magnetic moment \cite%
{Kiritsis:2002aj,Appelquist:2004mn,Giudice:2012ms,Omura:2015nja,Falkowski:2018dsl,Crivellin:2018qmi,Allanach:2015gkd,Padley:2015uma,Chen:2016dip,Raby:2017igl,Chiang:2017tai,Chen:2017hir,Megias:2017dzd,Davoudiasl:2018fbb,Liu:2018xkx,CarcamoHernandez:2019xkb,Nomura:2019btk,Kawamura:2019rth,Bauer:2019gfk,Botella:2018gzy,Han:2018znu,Wang:2018hnw,Dutta:2018fge,Badziak:2019gaf,Endo:2019bcj,Hiller:2019mou,CarcamoHernandez:2019ydc,CarcamoHernandez:2019lhv,Kawamura:2019hxp,Sabatta:2019nfg,Chen:2020tfr,CarcamoHernandez:2020pxw,Iguro:2019sly,Li:2020dbg,Arbelaez:2020rbq,Hiller:2020fbu,Jana:2020pxx,deJesus:2020ngn,deJesus:2020upp,Hati:2020fzp,Botella:2020xzf,Dorsner:2020aaz,Calibbi:2020emz,Dinh:2020pqn,Jana:2020joi,Chun:2020uzw,Chua:2020dya,Daikoku:2020nhr,Banerjee:2020zvi,Chen:2020jvl,Bigaran:2020jil,Kawamura:2020qxo,Endo:2020mev,Iguro:2020rby,Yin:2020afe,Chen:2021rnl,Athron:2021iuf,Arcadi:2021cwg,Das:2021zea,Yin:2021yqy,Yin:2021mls,Chiang:2021pma,Escribano:2021css,Zhang:2021gun,Yang:2021duj,Li:2021lnz,Hernandez:2021uxx,Hernandez:2021tii,Hernandez:2021mxo,CarcamoHernandez:2021iat}%
, anomaly not explained by the SM and recently confirmed by the Muon $g-2$
experiment at FERMILAB \cite{Abi:2021gix}.

Recently, three of us proposed a model of fermion mass generation, where the
fermion mass hierarchy arises from the sequential loop suppression, as
follows \cite{CarcamoHernandez:2016pdu}: 
\begin{eqnarray}
t\mbox{-quark} &\rightarrow &\mbox{{\it tree-level mass} from} \text{ Yukawa
couplings} ,  \label{eq:level-1} \\
b,c,\ \tau ,\mu &\rightarrow &\mbox{\it 1-loop mass};\ \mbox{tree-level}
\label{eq:level-2} \\
&&\hspace{20mm}\mbox{suppressed by a {\it symmetry}}.  \notag \\
s,u,d,\ e &\rightarrow &\mbox{\it 2-loop mass};\ \mbox{tree-level \& 1-loop}
\label{eq:level-3} \\
&&\hspace{20mm}\mbox{suppressed by a {\it symmetry}}.  \notag \\
\nu _{i} &\rightarrow &\mbox{\it n-loop mass}\ (n> 2);\ 
\mbox{tree-level \& lower
loops}  \label{eq:level-4} \\
&&\hspace{20mm}\mbox{suppressed by a {\it symmetry}}.  \notag \\[-3mm]
&&  \notag
\end{eqnarray}
with neutrino mass generated at 4-loop level ($n=4$). 
However, 
this model 
has a low cutoff scale, since it includes non-renormalizable Yukawa terms,
needed to implement the radiative mechanisms of the SM fermion mass
generation (\ref{eq:level-1})-(\ref{eq:level-4}). From the view-point of model building, it is much more preferable to have a renormalizable setup with a moderate amount of particle content and 
%
predicting a phenomenology beyond the SM
 within the reach of future experimental sensitivities. 
With this in mind, we propose here a renormalizable model implementing 
the sequential loop-suppression mechanism  (\ref{eq:level-1})-(\ref{eq:level-3}) with the light active neutrino masses appearing at three loop level ($n=3$).
This model
has a much
more economical field content compared to the similar renormalizable
models proposed in Refs.~\cite{CarcamoHernandez:2017cwi,CarcamoHernandez:2019cbd}.
%
%
%
For instance, whereas the scalar sector of the model of Ref.~\cite{CarcamoHernandez:2019cbd} has 2 $SU_{2L}$ scalar doublets, 7 complex
electrically neutral gauge singlet scalars and 5 electrically charged
singlet scalar fields, thus amounting to 32 scalar degrees of freedom, 
the model proposed here has three $SU_{2L}$ scalar doublets, 3 complex and 2
real electrically neutral singlet scalars, which corresponds to 20 scalar
degrees of freedom. Furthermore, the scalar sector of the model of Ref. \cite{CarcamoHernandez:2017cwi} has three $SU_{3L}$ scalar triplets, three complex
electrically neutral singlet scalars and four electrically charged singlet
scalar fields, thus amounting to 32 scalar degrees of freedom, which is much
larger than the number of scalar degrees of freedom of our current model.

Moreover, our model can also successfully accommodate the electron and muon
anomalous magnetic moments, the observed Dark Matter relic density, as well
the constraints arising from charged Lepton Flavor Violating (LFV) processes.

Let us emphasize the difference of our proposed model with respect to recent
publications based on radiative mass and hierarchy generation: In Ref.~\cite{Ibarra:2014fla} there is no mechanism to generate the masses of
the quarks of the first generation. In addition, the model described in \cite{Ibarra:2014fla} does not provide an explanation for the SM lepton mass
hierarchy. In Ref.~\cite{Altmannshofer:2014qha}, both the first and
second generation SM charged fermion masses are produced at one loop,
whereas here we generate the lightest SM charged fermion masses at two loop
level. Moreover, in~\cite{Altmannshofer:2014qha} the neutrinos remain
massless.

The paper is organized as follows. In section \ref{themodel} we outline the
proposed model. In section \ref{scalarpotential} we analyze the stability
and describe the electroweak symmetry breaking of the scalar potential of
the model. The scalar mass spectrum of the model is discussed in section \ref%
{scalarspectrum}. 
The implications of our model with respect to the SM fermion-mass hierarchy
is discussed in section \ref{fermionmasshierarchy}. In section \ref{LFV}
charged LFV decays as well as 
the constraints on the charged scalar masses are considered. The implications of
our model for the muon and electron anomalous magnetic moments are discussed
in section \ref{gminus2mu}. 
The prospects with respect to Dark Matter are analyzed in section \ref%
{darkmatter}. Our conclusions are given in section \ref{conclusions}.

\section{The Model}

\label{themodel}\vspace{-0.1cm} Before providing a complete model setup, let
us explain the 
motivations behind introducing extra scalars, fermions and symmetries needed
for a consistent implementation of the sequential loop suppression mechanism
for generating the SM fermion hierarchies.

Our strategy is to ban certain operators, by imposing appropriate symmetries
to ensure loop suppression, necessary to reproduce the observable hierarchy
of the SM fermion masses.

In our model the top quark mass arises at tree level from a renormalizable
Yukawa operator, with an order one Yukawa coupling, i.e. 
\begin{equation}
\overline{q}_{iL}\widetilde{\phi }u_{3R},\hspace{1cm}i=1,2,3 \;.
\end{equation}%
%
%
%
%
%
%
%
%
%
%
%
%
%
%
%
%
%
%
%
%
%
We denote the left-handed quarks by $q_{iL}$ and the right-handed up and
down quarks by $u_{iR}$ and $d_{iR}$, respectively, with $i=1,2,3$ the
family index. The SM like Higgs boson doublet is denoted by~$\phi$.

\begin{figure}[tbp]
\includegraphics[width = 1\textwidth]{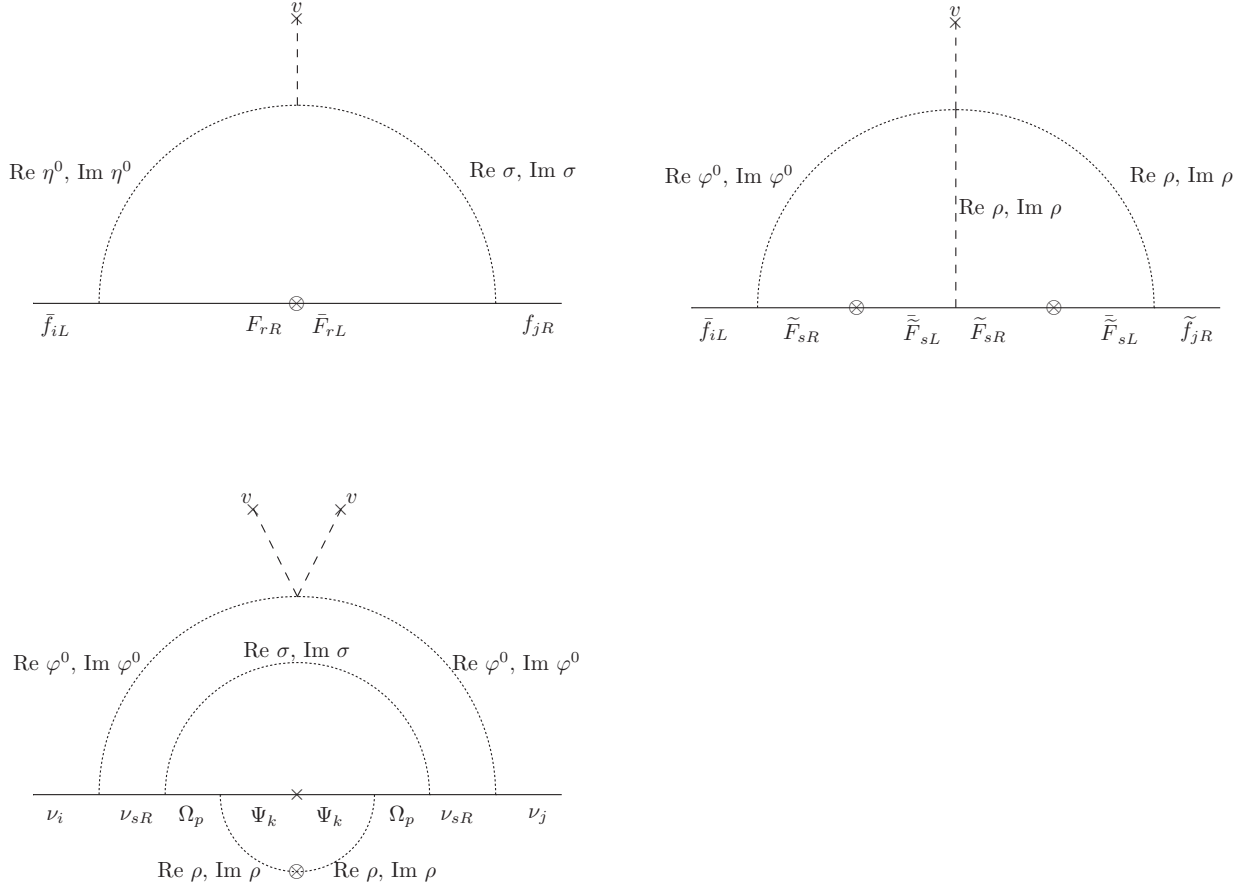}\vspace{-8.5cm}
\caption{Loop diagrams contributing to the fermion mass matrices.
Here $f_{iL}=u_{iL},d_{iL},e_{iL}$ ($i=1,2,3$), $%
f_{R}=u_{2R},d_{3R},l_{2R},l_{3R}$, $\widetilde{f}%
_{R}=u_{1R},d_{1R},d_{2R},l_{1R}$. The electroweak singlet charged exotic
fermions, see (\protect\ref{Fermions}), are denoted by $F_{rR}$, $F_{rL}$, $%
\widetilde{F}_{sR}$ and $\widetilde{F}_{sL}$, where $r=1$ for quarks, $r=2$
for charged leptons, $s=1$ for up type quarks and charged leptons, and $s=2$
for down type quarks and neutrinos. Furthermore, in the neutrino loop
diagram we have $p,k \in \{1,2\}$.}
\label{Loopdiagrams}
\end{figure}

To generate the bottom, charm, tau and muon masses at one loop level, it is
necessary to forbid the operators: 
\begin{eqnarray}
&&\overline{f}_{iL}Hf_{R},\hspace{1cm}f_{iL}=q_{iL},l_{iL},\hspace{1cm}%
f_{R}=u_{2R},d_{3R},l_{2R},l_{3R},  \notag \\
&&i=1,2,3,\hspace{1cm}{\text{with }}H=\left\{ 
\begin{array}{l}
\ \ \widetilde{\phi }\ \ \ \mbox{for}\ \ \ f_{R}=u_{2R}, \\ 
\\ 
\ \ \phi \ \ \mbox{for}\ \ \ f_{R}=d_{3R},l_{2R},l_{3R}.%
\end{array}%
\right. ,
\end{eqnarray}%
at tree level and to allow the following operators, crucial to close the one
loop level diagram of the upper left panel of Figure \ref{Loopdiagrams}: 
\begin{eqnarray}
&&\overline{f}_{iL}\Phi F_{rR},\hspace{1cm}\overline{F}_{rL}\sigma f_{R},%
\hspace{1cm}f_{iL}=q_{iL},l_{iL},\hspace{1cm}%
f_{R}=u_{2R},d_{3R},l_{2R},l_{3R},  \notag \\
&&i=1,2,3,\hspace{1cm}r=\left\{ 
\begin{array}{l}
\ \ 1\ \ \ \mbox{for quarks},\notag \\ 
\\ 
\ \ 2\ \ \ \mbox{for charged leptons}.%
\end{array}%
\right. ,\hspace{1cm}\Phi =\left\{ 
\begin{array}{l}
\ \ \widetilde{\eta }\ \ \ \mbox{for}\ \ \ f_{R}=u_{2R}, \\ 
\\ 
\ \ \eta \ \ \ \mbox{for}\ \ \ f_{R}=d_{3R},l_{2R},l_{3R}.%
\end{array}%
\right. , \\
&&A\left[ \left( \phi ^{\dagger }\eta \right) \sigma +h.c\right] ,\hspace{1cm%
}\hspace{1cm}\left( y_{F}\right) _{r}\overline{F}_{rL}\chi F_{rR}
\label{Op1}
\end{eqnarray}

This requires to add 
an unbroken $Z_{2}^{\left( 2\right) }$ symmetry as well as a spontaneously
broken $Z_{2}^{\left( 1\right) }$ symmetry. Under the spontaneously broken $%
Z_{2}^{\left( 1\right) }$ symmetry, all the right handed SM fermionic
fields, excepting $u_{3R}$ are charged. Under this $Z_{2}^{\left( 1\right) }$
symmetry, the singlet scalar field $\chi $ as well as the left-handed exotic
fermionic field $F_{rL}$ are charged. Furthermore, all SM fermionic fields
are neutral under the unbroken $Z_{2}^{\left( 2\right) }$ symmetry whereas
the left-handed and right-handed exotic fermionic fields $F_{rL}$ and $%
F_{rR} $ are charged under $Z_{2}^{\left( 2\right) }$. The inclusion of the
spontaneously broken $Z_{2}^{\left( 1\right) }$ and unbroken $Z_{2}^{\left(
2\right) }$ symmetries is crucial for the implementation of the radiative
seesaw mechanism that produces one-loop level masses for the bottom, charm,
tau and muon without invoking soft-breaking mass terms. Notice that the
fermionic sector 
is enlarged by 
electroweak charged exotic fermions $F_{r}$, where $r=1$ for quarks and $r=2$
for charged leptons, and that the Yukawa operators as well as the trilinear
scalar operator shown in Eq.~(\ref{Op1}) correspond to the three vertices of
the one loop level diagram in the upper left panel of Figure~\ref%
{Loopdiagrams}. Considering the simplest possibility, where such charged
exotic fermions $F_{r}$ are $SU_{2L}$ singlets, the scalar sector has to be
extended by the inclusion of an extra $SU_{2L}$ scalar doublet $\eta $ and
an electrically neutral electroweak gauge-singlet scalars $\sigma $ and $%
\chi $. The scalar fields $\eta $ and $\sigma $ are both charged under the
preserved $Z_{2}^{\left( 2\right) }$ symmetry, whereas the scalar $\chi $ is
neutral under this symmetry. The singlet scalar field $\chi $\ is needed to
provide masses to the charged exotic fermions $F_{r}$. This scalar field $%
\chi $ is assumed to be charged under the spontaneously broken $Z_{2}^{\left(
1\right) }$ symmetry. Furthermore, the Yukawa term $\left( y_{F}\right) _{r}%
\overline{F}_{rL}\chi F_{rR}$, which involves the electroweak charged exotic
fermions, must also be included as well, in order to close the one loop level
diagram of Figure \ref{Loopdiagrams}.

Besides that, small masses for the light SM charged fermions, i.e., the up,
down and strange quarks as well as the electron, 
are generated at two loop level. This implies to forbid the following
operators that would give rise to tree and one-loop-level masses for these
particles: 
\begin{eqnarray}
&&\overline{f}_{iL}Hf_{R},\hspace{0.5cm}f_{iL}=q_{iL},l_{iL},\hspace{0.5cm}%
f_{R}=u_{1R},d_{1R},d_{2R},l_{1R},\hspace{0.5cm}i=1,2,3,  \notag \\
&&H=\left\{ 
\begin{array}{l}
\ \ \widetilde{\phi }\ \ \ \mbox{for}\ \ \ f_{R}=u_{1R}, \\ 
\\ 
\ \ \phi \ \ \ \mbox{for}\ \ \ f_{R}=d_{1R},d_{2R},l_{1R}.%
\end{array}%
\right. ,  \notag \\
&&\overline{f}_{iL}\Phi F_{rR},\hspace{1cm}\overline{F}_{rL}\sigma f_{R},%
\hspace{1cm}f_{iL}=q_{iL},l_{iL},\hspace{1cm}%
f_{R}=u_{1R},d_{1R},d_{2R},l_{1R},  \notag \\
&&i=1,2,3,\hspace{1cm}r=\left\{ 
\begin{array}{l}
\ \ 1\ \ \ \mbox{for quarks}, \\ 
\\ 
\ \ 2\ \ \ \mbox{for charged leptons}.%
\end{array}%
\right. ,\hspace{1cm}\Phi =\left\{ 
\begin{array}{l}
\ \ \widetilde{\eta }\ \ \ \mbox{for}\ \ \ f_{R}=u_{2R}, \\ 
\\ 
\ \ \eta \ \ \ \mbox{for}\ \ \ f_{R}=d_{3R},l_{2R},l_{3R}.%
\end{array}%
\right. 
\end{eqnarray}%
%
%
%
%
%
%
%
%
%
%
%
%
%
%
%
%
%
%
%
%
%
However the following operators are required to provide two loop level
masses for the light SM charged fermions: 
\begin{eqnarray}
&&\overline{f}_{iL}\Xi \widetilde{F}_{sR},\hspace{1cm}\overline{\widetilde{F}%
}_{sL}\Delta \widetilde{F}_{sR}^{\prime },\hspace{1cm}\bar{\widetilde{F}}%
_{sL}^{\prime }\zeta \widetilde{F}_{sR}^{\prime },\hspace{1cm}\bar{%
\widetilde{F}}_{sL}^{\prime }\Theta \widetilde{f}_{R},  \notag \\
&&f_{iL}=q_{iL},l_{iL},\hspace{1cm}\Xi =\left\{ 
\begin{array}{l}
\ \ \widetilde{\varphi }\ \ \ \mbox{for}\ \ \ f_{R}=u_{1R},\notag \\ 
\\ 
\ \ \varphi \ \ \ \mbox{for}\ \ \ f_{R}=d_{1R},d_{2R},l_{1R}.%
\end{array}%
\right. ,\hspace{1cm}\widetilde{f}_{R}=u_{1R},d_{1R},d_{2R},l_{1R},\hspace{%
1cm}i=1,2,3  \notag \\
&&\Theta =\left\{ 
\begin{array}{l}
\ \ \xi ^{\ast }\ \ \mbox{for}\ \ \ f_{R}=u_{1R},d_{1R},d_{2R}\notag \\ 
\\ 
\ \ \xi \ \ \ \mbox{for}\ \ \ f_{R}=l_{1R}.%
\end{array}%
\right. ,\hspace{1cm}\Delta =\left\{ 
\begin{array}{l}
\ \ \rho ^{\ast }\ \ \mbox{for}\ \ \ \widetilde{F}_{sL}=\widetilde{T}_{L},%
\widetilde{B}_{sL}\notag \\ 
\\ 
\ \ \rho \ \ \ \mbox{for}\ \ \ \widetilde{F}_{sL}=\widetilde{E}_{L}.%
\end{array}%
\right. ,  \notag \\
&&\left[ \left( \phi ^{\dagger }\varphi \right) \rho \xi +h.c\right] ,%
\hspace{1cm}\hspace{1cm}\left( m_{\widetilde{F}}\right) _{s}\overline{%
\widetilde{F}}_{sL}\widetilde{F}_{sR},\hspace{1cm}\hspace{1cm}\left( y_{%
\widetilde{F}^{\prime }}\right) _{s}\bar{\widetilde{F}}_{sL}^{\prime }\zeta 
\widetilde{F}_{sR}^{\prime },  \notag \\
&&s=\left\{ 
\begin{array}{l}
\ \ 1\ \ \ \mbox{for up-type quarks and charged leptons}, \\ 
\\ 
\ \ 2\ \ \ \mbox{for down-type quarks}.%
\end{array}%
\right.   \label{Op2}
\end{eqnarray}%
Such operators are crucial to close the two-loop-level diagram of the upper
right panel of Figure~\ref{Loopdiagrams}. For this to happen, the fermion
sector 
is extended as well, by adding the electroweak charged exotic fermions $%
\widetilde{F}_{s}$, $\widetilde{F}_{s}^{\prime }$ where $s=1$ for up-type
quarks and charged exotic leptons and $s=2$ for down-type quarks. The
simplest choice is to assign 
these charged exotic fermions $\widetilde{F}_{s}$, $\widetilde{F}%
_{s}^{\prime }$ to $SU_{2L}$ singlets. Then, in order to build the Yukawa
interactions that determine three of the four vertices of the two loop level
diagram of Figure~\ref{Loopdiagrams}, we also need to add an extra $SU_{2L}$
scalar doublet $\varphi $ and another electrically neutral electroweak
gauge-singlet scalars $\rho $, $\xi $ and $\zeta $. The scalar fields $%
\varphi $, $\rho $ and $\xi $ are assumed to have complex charges under an
additional spontaneously broken $Z_{4}$ symmetry, whereas the scalar $\zeta $
has a real charge under this $Z_{4}$ symmetry. We further assume that the $%
Z_{4}$ symmetry is spontaneously broken down to a preserved $Z_{2}$
symmetry, which implies that the scalar fields $\rho $ and $\xi $ do not
acquire vacuum expectation values whereas the scalar $\zeta $ does.
Furthermore, in order to close the aforementioned two loop diagram,
one has to include the mass term $\left( m_{\widetilde{F}}\right) _{s}%
\overline{\widetilde{F}}_{sL}\widetilde{F}_{sR}$ and the Yukawa interaction $%
\left( y_{\widetilde{F}^{\prime }}\right) _{s}\bar{\widetilde{F}}%
_{sL}^{\prime }\zeta \widetilde{F}_{sR}^{\prime }$ involving the 
electroweakly charged exotic fermions. Notice that the Yukawa operators, as
well as the quartic scalar operator shown in Eq.~(\ref{Op2}), correspond to
the four vertices of the two loop level diagram of the upper right panel of
Figure~\ref{Loopdiagrams}.

In what regards the neutrino sector, we require that the light active
neutrino masses only appear at three-loop level. To this end, right-handed
Majorana-neutrinos have to be added in the fermionic spectrum. In addition,
one should prevent the appearance of tree, one and two-loop level masses for
the light active neutrinos. Generating light active neutrino masses at
three-loop level, as in the Feynman diagram of the bottom left panel of
Figure~\ref{Loopdiagrams}, requires the presence of the operators
\begin{equation}
\overline{l}_{jL}\widetilde{\varphi }\nu _{sR},\hspace{1cm}\overline{\nu
_{sR}^{C}}\sigma \Omega _{pR},\hspace{1cm}\overline{\nu _{sR}^{C}}\sigma
\Omega _{pR},\hspace{1cm}\overline{\Omega _{sR}^{C}}\rho \Psi _{pR},\hspace{%
1cm}\left( m_{\Psi }\right) _{sp}\Psi _{sR}\overline{\Psi _{pR}^{C}},\hspace{%
1cm}\left( \rho ^{2}\chi \zeta +h.c\right) ,
\end{equation}%
and forbidding:
\begin{equation}
\overline{l}_{jL}\phi \nu _{sR},\hspace{0.7cm}\hspace{0.7cm}m_{N}\nu _{sR}%
\overline{\nu _{sR}^{C}},\hspace{0.7cm}\hspace{0.7cm}\overline{l}_{jL}%
\widetilde{\eta }\nu _{sR},\hspace{0.7cm}\hspace{0.7cm}\left( m_{\Omega
}\right) _{sp}\Omega _{sR}\overline{\Omega _{pR}^{C}},
\end{equation}%
where $\nu _{sR}$, $\Omega _{sR}$ and $\Psi _{sR}$ ($s=1,2$) are gauge
singlet right-handed Majorana neutrinos. By an appropriate choice of charges
(shown below) under the aforementioned $Z_{2}^{\left( 1\right) }\times
Z_{2}^{\left( 2\right) }\times Z_{4}$ 
symmetry, the three-loop level radiative seesaw mechanism for light active
neutrinos can be implemented. 

With the aim of implementing the sequential loop suppression mechanism that
generates the pattern of SM fermion masses, we consider an extension of the
inert 3HDM, where the SM gauge symmetry is supplemented by a 
\mbox{$Z_{2}^{\left(1\right) }\times Z_{2}^{\left( 2\right) }\times Z_{4}$} discrete symmetry and
the scalar sector is extended to include five SM scalar singlets, i.e., $%
\sigma $, $\rho $, $\xi $, $\chi $ and $\zeta $. The reason to consider this
extra $Z_{2}^{\left( 1\right) }\times Z_{2}^{\left( 2\right) }\times Z_{4}$
discrete symmetry is that it is the smallest cyclic symmetry that allows us to realize the loop-suppression scenario (\ref{eq:level-1})-(\ref{eq:level-4}) with $n=3$
in a renormalizable 3HDM setup without invoking soft symmetry breaking.

The scalar sector of the model consists of three $SU_{2L}$ scalar doublets,
i.e., $\phi $, $\eta $, $\varphi $ and five scalar singlets $\sigma $, $\rho 
$, $\xi $, $\chi $ and $\zeta $, with the $Z_{2}^{\left( 1\right) }\times
Z_{2}^{\left( 2\right) }\times Z_{4}$ assignments: 
\begin{eqnarray}
\phi &\sim &\left( 1,1,1\right) ,\hspace{0.7cm}\eta \sim \left(
1,-1,1\right) ,\hspace{0.7cm}\varphi \sim \left( 1,-1,-1\right) ,\hspace{%
0.7cm}\sigma \sim \left( 1,-1,1\right) ,\hspace{0.7cm}\rho \sim \left(
1,-1,-i\right) ,  \notag \\
\xi &\sim &\left( 1,1,-i\right) ,\hspace{0.7cm}\chi \sim \left(
-1,1,1\right) ,\hspace{0.7cm}\zeta \sim \left( -1,1,-1\right)
\label{assignmentsscalars}
\end{eqnarray}%
We assume that the $Z_{2}^{\left( 2\right) }$ symmetry is 
unbroken whereas the $Z_{2}^{\left( 1\right) }$ and $Z_{4}$ symmetries are
spontaneously broken. We further assume that the $Z_{4}$ symmetry is
spontaneously broken down to a preserved $Z_{2}$ symmetry. These assumptions
imply that the scalar fields $\eta $, $\varphi $, $\sigma $, $\rho $, $\xi $%
, charged under the $Z_{2}^{\left( 2\right) }$ symmetry and (or) having
complex $Z_{4}$ charges, do not acquire vacuum expectation values. 
This conditions are inevitable in the present setup for implementing the scenario (\ref{eq:level-1})-(\ref{eq:level-4}).
%
Let us note that the $SU_{2L}$ scalar doublet $\phi $ is
the only scalar field that acquires a non-vanishing vacuum expectation value
(VEV) equal to about $246$ GeV and thus corresponds to the SM Higgs doublet.

A justification of the extension of the scalar sector of the model is
provided in the following. The $SU_{2L}$ inert scalar doublet $\eta $ as
well as the inert SM gauge singlet scalar $\sigma $ are introduced to
generate the one-loop level masses for the bottom, charm quarks, tau and
muon leptons. The scalar singlets $\chi $ and $\zeta $ are introduced to
provide masses to the charged exotic fermions. Moreover, the implementation
of the two-loop level radiative seesaw mechanisms, generating the up, down,
strange quark masses as well as the electron mass, requires to introduce an
extra $SU_{2L}$ inert scalar doublet, namely $\varphi $ and inert SM gauge
singlet scalars, i.e., $\rho $ and $\xi $%
. The particles $\varphi $ and $\rho $ are also crucial to give three-loop
level masses for the light active neutrinos. The three loop level neutrino
mass diagram is closed thanks to the gauge singlet scalars $\chi $ and $%
\zeta $. 

The fermion sector of the SM is extended by the $SU_{2L}$ singlet exotic
quarks $T$, $\tilde{T}$, $\widetilde{T}^{\prime }$, $B$, $\tilde{B}$, $%
\widetilde{B}^{\prime }$ and singlet leptons $E$, $\tilde{E}$, $\widetilde{E}%
^{\prime }$, $\nu _{s}$ ($s=1,2$), $\Omega $, $\Psi $ with electric charges $%
Q(T)=Q(\tilde{T})=2/3$, $Q(B)=Q(\tilde{B})=-1/3$, $Q(E)=-1$. The $%
Z_{2}^{\left( 1\right) }\times Z_{2}^{\left( 2\right) }\times Z_{4}$
assignments of the fermion sector are: 
\begin{eqnarray}
u_{1R} &\sim &\left( -1,1,-1\right) ,\hspace{1cm}u_{2R}\sim \left(
-1,1,1\right) ,\hspace{1cm}u_{3R}\sim \left( 1,1,1\right) ,  \notag \\
d_{1R} &\sim &\left( -1,1,-1\right) ,\hspace{1cm}d_{2R}\sim \left(
-1,1,-1\right) ,\hspace{1cm}d_{3R}\sim \left( -1,1,1\right) ,  \notag \\
l_{1R} &\sim &\left( -1,1,-i\right) ,\hspace{1cm}l_{2R}\sim \left(
-1,1,i\right) ,\hspace{0.5cm}l_{3R}\sim \left( -1,1,i\right) ,  \notag \\
q_{jL} &\sim &\left( 1,1,1\right) ,\hspace{1cm}l_{jL}\sim \left(
1,1,i\right) ,\hspace{1cm}j=1,2,3,  \notag \\
T_{L} &\sim &\left( -1,-1,1\right) ,\hspace{0.7cm}T_{R}\sim \left(
1,-1,1\right) ,\hspace{0.7cm}\widetilde{T}_{L}\sim \left( 1,-1,-1\right) ,%
\hspace{0.7cm}\widetilde{T}_{R}\sim \left( 1,-1,-1\right) ,\hspace{0.7cm} 
\notag \\
\widetilde{T}_{L}^{\prime } &\sim &\left( -1,1,-i\right) ,\hspace{0.7cm}%
\widetilde{T}_{R}^{\prime }\sim \left( 1,1,i\right) ,\hspace{0.7cm}B_{L}\sim
\left( -1,-1,1\right) ,\hspace{0.7cm}B_{R}\sim \left( 1,-1,1\right) ,  \notag
\\
B_{L} &\sim &\left( -1,-1,1\right) ,\hspace{1cm}B_{R}\sim \left(
1,-1,1\right) ,\hspace{1cm}\widetilde{B}_{sL}\sim \left( 1,-1,-1\right) ,%
\hspace{1cm}\widetilde{B}_{sR}\sim \left( 1,-1,-1\right) ,  \notag \\
\widetilde{B}_{L}^{\prime } &\sim &\left( -1,1,-i\right) ,\hspace{0.7cm}%
\widetilde{B}_{R}^{\prime }\sim \left( 1,1,i\right) ,\hspace{1cm}E_{sL}\sim
\left( -1,-1,i\right) ,\hspace{1cm}E_{sR}\sim \left( 1,-1,i\right) ,  \notag
\\
\widetilde{E}_{L} &\sim &\left( 1,-1,-i\right) ,\hspace{1cm}\widetilde{E}%
_{R}\sim \left( 1,-1,-i\right) ,\hspace{1cm}\widetilde{E}_{L}^{\prime }\sim
\left( -1,1,-1\right) ,\hspace{1cm}\widetilde{E}_{R}^{\prime }\sim \left(
1,1,1\right) ,  \notag \\
\nu _{sR} &\sim &\left( 1,-1,-i\right) ,\hspace{5mm}s=1,2,\hspace{1cm}\Omega
_{sR}\sim \left( 1,1,i\right) ,\hspace{1cm}\Psi _{sR}\sim \left(
1,-1,1\right) .  \label{Fermions}
\end{eqnarray}%
The quark, lepton and scalar assignments under $SU_{3c}\times SU_{2L}\times
U_{1Y}\times Z_{2}^{\left( 1\right) }\times Z_{2}^{\left( 2\right) }\times
Z_{4}$ are shown in Tables \ref{tab:quarks}, \ref{tab:leptons} and \ref%
{tab:scalars}, respectively.

\begin{table}[th]
\begin{tabular}{|c|c|c|c|c|c|c|c|c|c|c|c|c|c|c|c|c|c|c|c|}
\hline
& $q_{jL}$ & $u_{1R}$ & $u_{2R}$ & $u_{3R}$ & $d_{1R}$ & $d_{2R}$ & $d_{3R}$
& $T_{L}$ & $T_{R}$ & $\widetilde{T}_{L}$ & $\widetilde{T}_{R}$ & $%
\widetilde{T}_{L}^{\prime }$ & $\widetilde{T}_{R}^{\prime }$ & $B_{L}$ & $%
B_{R}$ & $\widetilde{B}_{sL}$ & $\widetilde{B}_{sR}$ & $\widetilde{B}%
_{sL}^{\prime }$ & $\widetilde{B}_{sR}^{\prime }$ \\ \hline
$SU_{3c}$ & $\mathbf{3}$ & $\mathbf{3}$ & $\mathbf{3}$ & $\mathbf{3}$ & $%
\mathbf{3}$ & $\mathbf{3}$ & $\mathbf{3}$ & $\mathbf{3}$ & $\mathbf{3}$ & $%
\mathbf{3}$ & $\mathbf{3}$ & $\mathbf{3}$ & $\mathbf{3}$ & $\mathbf{3}$ & $%
\mathbf{3}$ & $\mathbf{3}$ & $\mathbf{3}$ & $\mathbf{3}$ & $\mathbf{3}$ \\ 
\hline
$SU_{2L}$ & $\mathbf{2}$ & $\mathbf{1}$ & $\mathbf{1}$ & $\mathbf{1}$ & $%
\mathbf{1}$ & $\mathbf{1}$ & $\mathbf{1}$ & $\mathbf{1}$ & $\mathbf{1}$ & $%
\mathbf{1}$ & $\mathbf{1}$ & $\mathbf{1}$ & $\mathbf{1}$ & $\mathbf{1}$ & $%
\mathbf{1}$ & $\mathbf{1}$ & $\mathbf{1}$ & $\mathbf{1}$ & $\mathbf{1}$ \\ 
\hline
$U_{1Y}$ & $\frac{1}{6}$ & $\frac{2}{3}$ & $\frac{2}{3}$ & $\frac{2}{3}$ & $-%
\frac{1}{3}$ & -$\frac{1}{3}$ & $-\frac{1}{3}$ & $\frac{2}{3}$ & $\frac{2}{3}
$ & $\frac{2}{3}$ & $\frac{2}{3}$ & $\frac{2}{3}$ & $\frac{2}{3}$ & $-\frac{1%
}{3}$ & $-\frac{1}{3}$ & $-\frac{1}{3}$ & $-\frac{1}{3}$ & $-\frac{1}{3}$ & $%
-\frac{1}{3}$ \\ \hline
$Z_{2}^{\left( 1\right) }$ & $1$ & $-1$ & $-1$ & $1$ & $-1$ & $-1$ & $-1$ & $%
-1$ & $1$ & $1$ & $1$ & $-1$ & $1$ & $-1$ & $1$ & $1$ & $1$ & $-1$ & $1$ \\ 
\hline
$Z_{2}^{\left( 2\right) }$ & $1$ & $1$ & $1$ & $1$ & $1$ & $1$ & $1$ & $-1$
& $-1$ & $-1$ & $-1$ & $1$ & $1$ & $-1$ & $-1$ & $-1$ & $-1$ & $1$ & $1$ \\ 
\hline
$Z_{4}$ & $1$ & $-1$ & $1$ & $1$ & $-1$ & $-1$ & $1$ & $1$ & $1$ & $-1$ & $%
-1 $ & $-i$ & $i$ & $1$ & $1$ & $-1$ & $-1$ & $-i$ & $i$ \\ \hline
\end{tabular}%
\caption{Quark assignments under $SU_{3c}\times SU_{2L}\times U_{1Y}\times
Z_{2}^{\left( 1\right) }\times Z_{2}^{\left( 2\right) }\times Z_{4}$. Here $%
j=1,2,3$ and $s=1,2$.}
\label{tab:quarks}
\end{table}

\begin{table}[th]
\begin{tabular}{|c|c|c|c|c|c|c|c|c|c|c|c|c|c|}
\hline
& $l_{jL}$ & $l_{1R}$ & $l_{2R}$ & $l_{3R}$ & $E_{sL}$ & $E_{sR}$ & $%
\widetilde{E}_{L}$ & $\widetilde{E}_{R}$ & $\widetilde{E}_{L}^{\prime }$ & $%
\widetilde{E}_{R}^{\prime }$ & $\nu _{sR}$ & $\Omega _{sR}$ & $\Psi _{sR}$
\\ \hline
$SU_{3c}$ & $\mathbf{1}$ & $\mathbf{1}$ & $\mathbf{1}$ & $\mathbf{1}$ & $%
\mathbf{1}$ & $\mathbf{1}$ & $\mathbf{1}$ & $\mathbf{1}$ & $1$ & $1$ & $%
\mathbf{1}$ & $\mathbf{1}$ & $\mathbf{1}$ \\ \hline
$SU_{2L}$ & $\mathbf{2}$ & $\mathbf{1}$ & $\mathbf{1}$ & $\mathbf{1}$ & $%
\mathbf{1}$ & $\mathbf{1}$ & $\mathbf{1}$ & $\mathbf{1}$ & $\mathbf{1}$ & $%
\mathbf{1}$ & $\mathbf{1}$ & $\mathbf{1}$ & $\mathbf{1}$ \\ \hline
$U_{1Y}$ & $-\frac{1}{2}$ & $-1$ & $-1$ & $-1$ & $-1$ & $-1$ & $-1$ & $-1$ & 
$-1$ & $-1$ & $0$ & $0$ & $0$ \\ \hline
$Z_{2}^{\left( 1\right) }$ & $1$ & $-1$ & $-1$ & $-1$ & $-1$ & $1$ & $1$ & $%
1 $ & $-1$ & $1$ & $1$ & $1$ & $1$ \\ \hline
$Z_{2}^{\left( 2\right) }$ & $1$ & $1$ & $1$ & $1$ & $-1$ & $-1$ & $-1$ & $%
-1 $ & $1$ & $1$ & $-1$ & $1$ & $-1$ \\ \hline
$Z_{4}$ & $i$ & $-i$ & $i$ & $i$ & $i$ & $i$ & $-i$ & $-i$ & $-1$ & $1$ & $%
-i $ & $i$ & $1$ \\ \hline
\end{tabular}%
\caption{Lepton assignments under $SU_{3c}\times SU_{2L}\times U_{1Y}\times
Z_{2}^{\left( 1\right) }\times Z_{2}^{\left( 2\right) }\times Z_{4}$. Here $%
j=1,2,3$ and $s=1,2$.}
\label{tab:leptons}
\end{table}

\begin{table}[th]
\begin{tabular}{|c|c|c|c|c|c|c|c|c|}
\hline
& $\phi $ & $\eta $ & $\varphi $ & $\sigma $ & $\rho $ & $\xi $ & $\chi $ & $%
\zeta $ \\ \hline
$SU_{3c}$ & $\mathbf{1}$ & $\mathbf{1}$ & $\mathbf{1}$ & $\mathbf{1}$ & $%
\mathbf{1}$ & $\mathbf{1}$ & $\mathbf{1}$ & $\mathbf{1}$ \\ \hline
$SU_{2L}$ & $\mathbf{2}$ & $\mathbf{2}$ & $\mathbf{2}$ & $\mathbf{1}$ & $%
\mathbf{1}$ & $\mathbf{1}$ & $\mathbf{1}$ & $\mathbf{1}$ \\ \hline
$U_{1Y}$ & $\frac{1}{2}$ & $\frac{1}{2}$ & $\frac{1}{2}$ & $0$ & $0$ & $0$ & 
$0$ & $0$ \\ \hline
$Z_{2}^{(1)}$ & $1$ & $1$ & $1$ & $1$ & $1$ & $1$ & $-1$ & $-1$ \\ \hline
$Z_{2}^{(2)}$ & $1$ & $-1$ & $-1$ & $-1$ & $-1$ & $1$ & $1$ & $1$ \\ \hline
$Z_{4}$ & $1$ & $1$ & $-1$ & $1$ & $-i$ & $-i$ & $1$ & $-1$ \\ \hline
\end{tabular}%
\caption{Scalar assignments under $SU_{3c}\times SU_{2L}\times U_{1Y}\times
Z_{2}^{(1)}\times Z_{2}^{(2)}\times Z_{4}$.}
\label{tab:scalars}
\end{table}

Now, let us justify 
the exotic fermion content of our model.
The gauge-singlet neutral leptons $\nu _{s}$, $\Omega _{s}$, $%
\Psi _{s}$ ($s=1,2$) are introduced to generate the three-loop level masses
for two light active neutrinos. Let us note that the neutrino oscillation
experimental data requires to have at least two light massive active
neutrinos \cite{Fogli:1998au}. Furthermore, note that the $SU_{2L}$ singlet
exotic quarks $T$, $\tilde{T}$, $\widetilde{T}^{\prime }$, $B$, $\tilde{B}$, 
$\widetilde{B}^{\prime }$and singlet leptons $E_{s}$ ($s=1,2$), $\tilde{E}$, 
$\widetilde{E}^{\prime }$ introduced in our model, correspond to the minimal
amount of charged exotic fermion content needed to yield one-loop level
masses for the bottom, charm quarks, tau and muon leptons, as well as two-
loop level masses for the light up, down, strange quarks and the electron,
without including soft-breaking mass terms.

With the specified particle content, we have the following quark, charged
lepton and neutrino Yukawa terms invariant under the $Z_{2}^{\left( 1\right)
}\times Z_{2}^{\left( 2\right) }\times Z_{4}$ discrete symmetry 
\begin{eqnarray}
-\mathcal{L}_{\text{Y}}^{\left( U\right) } &=&\sum_{j=1}^{3}y_{j}^{\left(
u\right) }\overline{q}_{jL}\widetilde{\varphi }\widetilde{T}_{R}+x^{\left(
u\right) }\bar{\widetilde{T}}_{L}^{\prime }\xi ^{\ast
}u_{1R}+\sum_{j=1}^{3}z_{j}^{\left( u\right) }\overline{q}_{jL}\widetilde{%
\eta }T_{R}+w^{\left( u\right) }\overline{T}_{L}\sigma u_{2R}  \notag \\
&+&\sum_{j=1}^{3}y_{j3}^{\left( u\right) }\overline{q}_{jL}\widetilde{\phi }%
u_{3R}+y_{T}\overline{T}_{L}\chi T_{R}+\widetilde{m}_{T}\overline{\widetilde{%
T}}_{L}\widetilde{T}_{R}+y_{\widetilde{T}^{\prime }}\bar{\widetilde{T}}%
_{L}^{\prime }\zeta \widetilde{T}_{R}^{\prime }+z_{\widetilde{T}^{\prime }}%
\overline{\widetilde{T}}_{L}\rho ^{\ast }\widetilde{T}_{R}^{\prime }+h.c,
\label{lyu}
\end{eqnarray}%
\begin{eqnarray}
-\mathcal{L}_{\text{Y}}^{\left( D\right) }
&=&\sum_{j=1}^{3}\sum_{s=1}^{2}y_{js}^{\left( d\right) }\overline{q}%
_{jL}\varphi \widetilde{B}_{sR}+\sum_{s=1}^{2}\sum_{k=1}^{2}x_{sk}^{\left(
d\right) }\bar{\widetilde{B}}_{sL}^{\prime }\xi ^{\ast
}d_{kR}+\sum_{j=1}^{3}z_{j}^{\left( d\right) }\overline{q}_{jL}\eta
B_{R}+w^{\left( d\right) }\overline{B}_{L}\sigma d_{3R}  \notag \\
&&+y_{B}\overline{B}_{L}\chi B_{R}+\sum_{s=1}^{2}\widetilde{m}_{B_{s}}%
\overline{\widetilde{B}}_{sL}\widetilde{B}_{sR}+\sum_{s=1}^{2}\left( y_{%
\widetilde{B}^{\prime }}\right) _{s}\bar{\widetilde{B}}_{sL}^{\prime }\zeta 
\widetilde{B}_{sR}^{\prime }+\sum_{s=1}^{2}\left( x_{\widetilde{B}^{\prime
}}\right) _{s}\overline{\widetilde{B}}_{sL}\rho ^{\ast }\widetilde{B}%
_{sR}^{\prime }+h.c,  \label{lyd}
\end{eqnarray}%
\begin{eqnarray}
-\mathcal{L}_{\text{Y}}^{\left( l\right) } &=&\sum_{j=1}^{3}y_{j}^{\left(
l\right) }\overline{l}_{jL}\varphi \widetilde{E}_{R}+x_{1}^{\left( l\right) }%
\bar{\widetilde{E}}_{L}^{\prime }\xi
l_{1R}+\sum_{j=1}^{3}\sum_{s=1}^{2}y_{js}^{\left( l\right) }\overline{l}%
_{jL}\eta E_{sR}+\sum_{s=1}^{2}\sum_{k=2}^{3}x_{sk}^{\left( l\right) }%
\overline{E}_{sL}\sigma l_{kR}  \notag \\
&&+\sum_{s=1}^{2}y_{E_{s}}\overline{E}_{sL}\chi E_{sR}+\widetilde{m}_{E}%
\overline{\widetilde{E}}_{L}\widetilde{E}_{R}+y_{\widetilde{E}^{\prime }}%
\bar{\widetilde{E}}_{L}^{\prime }\zeta \widetilde{E}_{R}^{\prime }+z_{%
\widetilde{E}^{\prime }}\overline{\widetilde{E}}_{L}\rho \widetilde{E}%
_{R}^{\prime }+h.c,  \label{lyl}
\end{eqnarray}%
\begin{equation}
-\mathcal{L}_{\text{Y}}^{\left( \nu \right)
}=\sum_{j=1}^{3}\sum_{s=1}^{2}y_{js}^{\left( \nu \right) }\overline{l}_{jL}%
\widetilde{\varphi }\nu _{sR}+\sum_{s=1}^{2}\sum_{p=1}^{2}y_{sp}^{\left( \nu
\right) }\overline{\nu _{sR}^{C}}\sigma \Omega
_{pR}+\sum_{s=1}^{2}\sum_{p=1}^{2}y_{sp}^{\left( \Omega \right) }\overline{%
\Omega _{sR}^{C}}\rho \Psi _{pR}+\sum_{s=1}^{2}\sum_{p=1}^{2}\left( m_{\Psi
}\right) _{sp}\Psi _{sR}\overline{\Psi _{pR}^{C}}+h.c.  \label{lnu}
\end{equation}%
%
%
%
%
%
%
%
%
%
%
%
%
%
%
%
%
After electroweak gauge-symmetry breaking, the above-given Yukawa
interactions yield the SM fermion masses via sequential loop suppression.
Furthermore, the non SM-like scalars (excepting the scalar singlets $\chi $
and $\zeta $) are not allowed to acquire VEVs for the following reasons: 
Firstly, in this way we avoid to generate %
tree-level masses for the SM fermions lighter than the top quark. 
Secondly, we open the possibility to have 
stable scalar Dark Matter candidates. 
Eventually we also avoid to encounter tree level Flavor Changing Neutral
Currents (FCNCs).

\section{Stability and electroweak symmetry breaking of the Higgs potential}

\label{scalarpotential} The renormalizable Higgs potential, invariant under
the symmetries of the model, has the form:

\begin{eqnarray}
V &=&\mu _{1}^{2}\left( \phi ^{\dagger }\phi \right) +\mu _{2}^{2}\left(
\eta ^{\dagger }\eta \right) +\mu _{3}^{2}\left( \varphi ^{\dagger }\varphi
\right) +\mu _{4}^{2}\left\vert \sigma \right\vert ^{2}+\left[ \mu
_{5}^{2}\sigma ^{2}+h.c\right] +\mu _{6}^{2}\left\vert \rho \right\vert
^{2}+\mu _{7}^{2}\left\vert \xi \right\vert ^{2}+\mu _{8}^{2}\chi ^{2}+\mu
_{9}^{2}\zeta ^{2}  \notag \\
&&+\lambda _{1}\left( \phi ^{\dagger }\phi \right) ^{2}+\lambda _{2}\left(
\eta ^{\dagger }\eta \right) ^{2}+\lambda _{3}\left( \varphi ^{\dagger
}\varphi \right) ^{2}+\lambda _{4}\left( \phi ^{\dagger }\phi \right) \left(
\eta ^{\dagger }\eta \right) +\lambda _{5}\left( \phi ^{\dagger }\phi
\right) \left( \varphi ^{\dagger }\varphi \right)  \notag \\
&&+\lambda _{6}\left( \eta ^{\dagger }\eta \right) \left( \varphi ^{\dagger
}\varphi \right) +\lambda _{7}\left( \phi ^{\dagger }\eta \right) \left(
\eta ^{\dagger }\phi \right) +\lambda _{8}\left( \phi ^{\dagger }\varphi
\right) \left( \varphi ^{\dagger }\phi \right) +\lambda _{9}\left( \eta
^{\dagger }\varphi \right) \left( \varphi ^{\dagger }\eta \right)  \notag \\
&&+\left[ \frac{\lambda _{10}}{2}\left( \phi ^{\dagger }\eta \right) ^{2}+h.c%
\right] +\left[ \frac{\lambda _{11}}{2}\left( \phi ^{\dagger }\varphi
\right) ^{2}+h.c\right] +\left[ \frac{\lambda _{12}}{2}\left( \eta ^{\dagger
}\varphi \right) ^{2}+h.c\right]  \notag \\
&&+\kappa _{1}\left\vert \sigma \right\vert ^{4}+\kappa _{2}\left\vert \rho
\right\vert ^{4}+\kappa _{3}\left\vert \xi \right\vert ^{4}+\kappa _{4}\chi
^{4}+\kappa _{5}\zeta ^{4}+\kappa _{6}\left\vert \sigma \right\vert
^{2}\left\vert \rho \right\vert ^{2}+\kappa _{7}\left\vert \sigma
\right\vert ^{2}\left\vert \xi \right\vert ^{2}+\kappa _{8}\left\vert \rho
\right\vert ^{2}\left\vert \xi \right\vert ^{2}  \notag \\
&&+\kappa _{9}\left\vert \rho \right\vert ^{2}\left\vert \xi \right\vert
^{2}+\kappa _{10}\chi ^{2}\zeta ^{2}+\kappa _{11}\left\vert \sigma
\right\vert ^{2}\chi ^{2}+\kappa _{12}\left\vert \rho \right\vert ^{2}\chi
^{2}+\kappa _{13}\left\vert \xi \right\vert ^{2}\chi ^{2}+\kappa
_{14}\left\vert \sigma \right\vert ^{2}\zeta ^{2}+\kappa _{15}\left\vert
\rho \right\vert ^{2}\zeta ^{2}  \notag \\
&&+\kappa _{16}\left\vert \xi \right\vert ^{2}\zeta ^{2}+\kappa _{17}\left(
\rho ^{2}\zeta \chi +h.c\right) +\kappa _{18}\left( \xi ^{2}\zeta \chi
+h.c\right) +\alpha _{1}\left( \phi ^{\dagger }\phi \right) \left\vert
\sigma \right\vert ^{2}+\left[ \alpha _{2}\sigma ^{2}+h.c\right] \left( \phi
^{\dagger }\phi \right)  \notag \\
&&+\alpha _{3}\left( \phi ^{\dagger }\phi \right) \left\vert \rho
\right\vert ^{2}+\alpha _{4}\left( \phi ^{\dagger }\phi \right) \left\vert
\xi \right\vert ^{2}+\alpha _{5}\left( \eta ^{\dagger }\eta \right)
\left\vert \sigma \right\vert ^{2}+\left[ \alpha _{6}\sigma ^{2}+h.c\right]
\left( \eta ^{\dagger }\eta \right) +\alpha _{7}\left( \eta ^{\dagger }\eta
\right) \left\vert \rho \right\vert ^{2}  \notag \\
&&+\alpha _{8}\left( \eta ^{\dagger }\eta \right) \left\vert \xi \right\vert
^{2}+\alpha _{9}\left( \varphi ^{\dagger }\varphi \right) \left\vert \sigma
\right\vert ^{2}+\left[ \alpha _{10}\sigma ^{2}+h.c\right] \left( \varphi
^{\dagger }\varphi \right) +\alpha _{11}\left( \varphi ^{\dagger }\varphi
\right) \left\vert \rho \right\vert ^{2}+\alpha _{12}\left( \varphi
^{\dagger }\varphi \right) \left\vert \xi \right\vert ^{2}  \notag \\
&&+\alpha _{13}\left( \phi ^{\dagger }\phi \right) \chi ^{2}+\alpha
_{14}\left( \phi ^{\dagger }\phi \right) \zeta ^{2}+\alpha _{15}\left( \eta
^{\dagger }\eta \right) \chi ^{2}+\alpha _{16}\left( \eta ^{\dagger }\eta
\right) \zeta ^{2}+\alpha _{17}\left( \varphi ^{\dagger }\varphi \right)
\chi ^{2}+\alpha _{18}\left( \varphi ^{\dagger }\varphi \right) \zeta ^{2} 
\notag \\
&&+\left[ A\left( \phi ^{\dagger }\eta \right) \sigma +h.c\right] +\left[
B\left( \rho ^{\ast }\xi \right) \sigma +h.c\right] +\gamma \left[ \left(
\phi ^{\dagger }\eta \right) \rho ^{\ast }\xi +h.c\right] +\varkappa \left[
\left( \phi ^{\dagger }\varphi \right) \rho \xi +h.c\right] .  \label{V}
\end{eqnarray}%
\label{expansion} The scalar fields can be written as

\begin{eqnarray}
\phi &=&\left( 
\begin{array}{c}
\phi ^{+} \\ 
\frac{1}{\sqrt{2}}\left( v+\phi _{R}^{0}+i\phi _{I}^{0}\right)%
\end{array}%
\right) ,\qquad \eta =\left( 
\begin{array}{c}
\eta ^{+} \\ 
\frac{1}{\sqrt{2}}\left( \eta _{R}^{0}+i\eta _{I}^{0}\right)%
\end{array}%
\right) ,\qquad \varphi =\left( 
\begin{array}{c}
\varphi ^{+} \\ 
\frac{1}{\sqrt{2}}\left( \varphi _{R}^{0}+i\varphi _{I}^{0}\right)%
\end{array}%
\right) , \\
\sigma &=&\frac{1}{\sqrt{2}}\left( \sigma _{R}+i\sigma _{I}\right) ,\qquad
\rho =\frac{1}{\sqrt{2}}\left( \rho _{R}+i\rho _{I}\right) ,\qquad \xi =%
\frac{1}{\sqrt{2}}\left( \xi _{R}+i\xi _{I}\right) ,\qquad \chi =v_{\chi }+%
\widetilde{\chi },\qquad \zeta =v_{\zeta }+\widetilde{\zeta }.
\end{eqnarray}%
From the condition to have a vanishing gradient of the potential with the
neutral component of the doublet $\phi $ getting a VEV $v/\sqrt{2}$ as well
as the scalar singlets $\chi $, $\zeta $ acquiring VEVs $v_{\chi }$ and $%
v_{\zeta }$, respectively, whereas all other VEVs are vanishing, 
we find the constraints on the potential parameters: 
\begin{eqnarray}
\mu _{1}^{2} &=&-\lambda _{1}v^{2}-\alpha _{13}v_{\chi }^{2}-\alpha
_{14}v_{\zeta }^{2},  \label{tadpole} \\
\mu _{8}^{2} &=&-2\kappa _{4}v_{\chi }^{2}-\kappa _{10}v_{\zeta }^{2}-\alpha
_{13}\frac{v^{2}}{2}, \\
\mu _{9}^{2} &=&-2\kappa _{5}v_{\zeta }^{2}-\kappa _{10}v_{\chi }^{2}-\alpha
_{14}\frac{v^{2}}{2},
\end{eqnarray}

From the symmetry of the potential \eqref{V} we find that for positive
quartic parameters $\lambda _{1}$, $\lambda _{2}$, $\lambda _{3}$, as well
as $\kappa _{1}$, $\kappa _{2}$, $\kappa _{3}$, $\kappa _{4}$, $\kappa _{5}$%
, the potential is bounded from below, that is, it is stable. However, we
also have to ensure that it provides the experimentally acceptable
electroweak symmetry breaking of $SU(2)_{L}\times U(1)_{Y}\rightarrow U(1)_{%
\text{em}}$ and gives the correct VEV of about $v\approx 246$ GeV.

Let us emphasize that in general it is not sufficient to check that the
potential has a vanishing gradient, leading to the condition \eqref{tadpole}.

In particular, the corresponding local stationary point can correspond to a
saddle point or maximum, and not a minimum. Moreover, there can be deeper
stationary points. A systematic approach to find the global minimum for any
3HDM has been presented in \cite{Maniatis:2014oza}. The case of two
Higgs-boson doublets accompanied by an arbitrary number of Higgs-boson
singlets has been also studied \cite{Maniatis:2006jd}. For the potential
considered here we have, in addition to the three Higgs-boson doublet fields 
$\phi $, $\eta $, $\varphi $, also three complex singlet fields, $\sigma $, $%
\rho $ and $\xi $ and two real scalars $\chi $ and $\zeta $. We adopt the
formalism of three doublets presented in \cite{Maniatis:2014oza} to the case
of additional Higgs-singlet fields that we have here.

The essential step is to introduce \emph{bilinears} for the Higgs-boson
doublets \cite{Nishi:2006tg, Maniatis:2006fs} and decompose the complex
Higgs singlets into its real and imaginary parts. First, all gauge-invariant
scalar products of the three doublet fields $\phi$, $\eta$, $\varphi$ are
arranged in a matrix, 
\begin{equation}
\underline{K}=\left( 
\begin{array}{ccc}
\phi ^{\dagger }\phi & \eta ^{\dagger }\phi & \varphi ^{\dagger }\phi \\ 
\phi ^{\dagger }\eta & \eta ^{\dagger }\eta & \varphi ^{\dagger }\eta \\ 
\phi ^{\dagger }\varphi & \eta ^{\dagger }\varphi & \varphi ^{\dagger
}\varphi%
\end{array}%
\right) .
\end{equation}%
This matrix can be expressed in a basis of matrices $\lambda_\alpha$ ($%
\alpha = 0, 1, \ldots,8$), where $\lambda_0 = \sqrt{\frac{2}{3}} \mathbbm{1}%
_3$ is the conveniently scaled identity matrix and $\lambda_a$ ($%
a=1,\ldots,8 $) are the Gell-Mann matrices. In this basis we can write 
\begin{equation}
\underline{K} = \frac{1}{2} \sum_{\alpha=0}^8 K_\alpha \lambda_\alpha.
\end{equation}
The real coefficients, called \emph{bilinears} $K_\alpha$, are obtained from 
\begin{equation}
K_\alpha = K_\alpha^* = \trace (\underline{K} \lambda_\alpha), \qquad \alpha
= 0, \ldots, 8.
\end{equation}
We can invert this relation and express the gauge-invariant scalar products
of the doublets, which appear in the potential, in terms of the bilinears: 
\begin{equation}  \label{bilinear}
\begin{split}
\phi ^{\dagger }\phi &=\frac{K_{0}}{\sqrt{6}}+\frac{K_{3}}{2}+\frac{K_{8}}{2%
\sqrt{3}}, \qquad \phi ^{\dagger }\eta =\frac{1}{2}\left(
K_{1}+iK_{2}\right), \qquad \phi ^{\dagger}\varphi =\frac{1}{2}\left(
K_{4}+iK_{5}\right), \\
\eta ^{\dagger }\eta &= \frac{K_{0}}{\sqrt{6}}-\frac{K_{3}}{2}+\frac{K_{8}}{2%
\sqrt{3}}, \qquad \eta ^{\dagger }\varphi =\frac{1}{2}\left(
K_{6}+iK_{7}\right) , \qquad \varphi ^{\dagger}\varphi =\frac{K_{0}}{\sqrt{6}%
}-\frac{K_{8}}{\sqrt{3}}.
\end{split}%
\end{equation}

Further, we decompose the complex singlets into its real and imaginary
parts, 
\begin{equation}
\sigma =\frac{1}{\sqrt{2}}\left( \sigma _{R}+i\sigma _{I}\right) ,\qquad
\rho =\frac{1}{\sqrt{2}}\left( \rho _{R}+i\rho _{I}\right) ,\qquad \xi =%
\frac{1}{\sqrt{2}}\left( \xi _{R}+i\xi _{I}\right) .  \label{singdec}
\end{equation}

With the replacements \eqref{bilinear} and \eqref{singdec}, the potential
can be written in terms of the bilinears as well as the real and imaginary
parts of the singlets, $V(K_{0},\ldots ,K_{8},\sigma _{R},\sigma _{I},\rho
_{R},\rho _{I},\xi _{R},\xi _{I},\chi ,\zeta )$.

All gauge degrees of freedom are systematically avoided and all fields and
parameters are real in this form.

We now look for all stationary points of the potential, in order to find the
global minimum, or in the degenerate case, the global minima. For a stable
potential, the global minimum is given by the deepest stationary point. We
now classify the stationary points with respect to the rank of the matrix $%
\underline{K}$. Any stationary point with rank~2 of $\underline{K}$
corresponds to a fully broken electroweak symmetry, rank~0 to an unbroken
electroweak symmetry, and rank~1 to a physically acceptable breaking of $%
SU(2)_{L}\times U(1)_{Y}\rightarrow U(1)_{\text{em}}$. The rank conditions
result in different sets of polynomial equations. Explicitly, the set of
equations corresponding to the rank~2 are, 
\begin{eqnarray}
&&\nabla _{K_{0},\ldots ,K_{8},\sigma _{R},\sigma _{I},\rho _{R},\rho
_{I},\xi _{R},\xi _{I},\chi ,\zeta }\bigg[V(K_{0},\ldots ,K_{8},\sigma
_{R},\sigma _{I},\rho _{R},\rho _{I},\xi _{R},\xi _{I},\chi ,\zeta )-u\;\det
(\underline{K})\bigg]=0,  \notag \\
&&2K_{0}^{2}-\sum_{a=1}^{8}K_{a}K_{a}>0,  \notag \\
&&\det (\underline{K})=0,  \notag \\
&&K_{0}>0.  \label{set2}
\end{eqnarray}%
Here $u$ denotes a Lagrange multiplier.

For the solutions with rank~0 of $\underline{K}$, we set all bilinears to
zero and look for the stationary points of the corresponding potential, that
is, solutions of the set of equations, 
\begin{equation}
\nabla _{\sigma _{R},\sigma _{I},\rho _{R},\rho _{I},\xi _{R},\xi _{I},\chi
,\zeta }V(K_{0}=0,\ldots ,K_{8}=0,\sigma _{R},\sigma _{I},\rho _{R},\rho
_{I},\xi _{R},\xi _{I},\chi ,\zeta )=0.  \label{set0}
\end{equation}

With respect to rank~1 solutions of $\underline{K}$, we can parametrize the
matrix $\underline{K}$ in terms of the three-component complex vector $w=%
\begin{pmatrix}
w_{1},w_{2},w_{3}%
\end{pmatrix}%
^{\mathrm{T}}$, 
\begin{equation}
\underline{K}=K_{0}\sqrt{\frac{3}{2}}\boldsymbol{w}\boldsymbol{w}^{\dagger },
\end{equation}%
getting for the bilinears 
\begin{equation}
K_{\alpha }(K_{0},\boldsymbol{w}^{\dagger },\boldsymbol{w})=K_{0}\sqrt{\frac{%
3}{2}}\boldsymbol{w}^{\dagger }\lambda _{\alpha }\boldsymbol{w},\qquad
\alpha =0,\ldots ,8.
\end{equation}%
The potential can now be written as $V(K_{0},\boldsymbol{w}^{\dagger },%
\boldsymbol{w},\sigma _{R},\sigma _{I},\rho _{R},\rho _{I},\xi _{R},\xi
_{I},\chi ,\zeta )$ and the corresponding set of polynomial equations reads 
\begin{equation}
\begin{split}
& \nabla _{K_{0},w_{1},w_{2},w_{3},\sigma _{R},\sigma _{I},\rho _{R},\rho
_{I},\xi _{R},\xi _{I},\chi ,\zeta }\left[ V(K_{0},\boldsymbol{w}^{\dagger },%
\boldsymbol{w},\sigma _{R},\sigma _{I},\rho _{R},\rho _{I},\xi _{R},\xi
_{I},\chi ,\zeta )-u(\boldsymbol{w}^{\dagger }\boldsymbol{w}-1)\right] =0, \\
& \boldsymbol{w}^{\dagger }\boldsymbol{w}-1=0, \\
& K_{0}>0,
\end{split}
\label{set1}
\end{equation}%
where $u$ again denotes a Lagrange mulitplier. We solve the three sets of
equations \eqref{set2}, \eqref{set0}, \eqref{set1} and the solution with the
lowest potential value is (are) the global minimum (minima). A solution is
only physically acceptable if it originates from the set \eqref{set1},
corresponding to the observed electroweak symmetry-breaking. In addition, we
have to check that the vacuum gives the observed VEV 
$v$. Numerically, we accept solutions which provide a vacuum-expectation
value in the range $245\text{ GeV}<v<247\text{ GeV}$. The sets of equations
can be solved via homotopy continuation; see for instance~\cite%
{Maniatis:2012ex}. The homotopy continuation algorithms can be found
implemented in the open-source software package PHCpack~\cite{phcpack}. 
We have numerically checked that there is large parameter space available
fulfilling the stationarity and stability conditions of the potential. Also
the potential can provide sufficiently heavy scalars, apart from the SM-like
Higgs boson, in accordance with the experimental constraints.

\section{The Higgs mass spectrum}

\label{scalarspectrum}

Here we restrict ourselves 
to real parameters of the potential~\eqref{V}, that is, in particular we
consider a CP conserving scalar sector. Then, we find that the spectrum of
the physical CP even neutral scalars is composed of the~$126$~GeV SM-like
Higgs boson, i.e $h$, two heavy CP even scalars $H_{1}$ and $H_{2}$ as well
as the inert scalars transforming non-trivially under the $Z_{2}^{\left(
2\right) }$ symmetry and (or) having complex $Z_{4}$ charges, namely $%
\varphi _{R}^{0}$, $\rho _{R}$, $\xi _{R}$, $S_{1}$ and $S_{2}.$ For the
sake of simplicity, we consider the scenario of the decoupling limit which
is motivated by the experimental fact that the couplings of the $126$~GeV
SM-like Higgs boson are very close to the SM expectation. In the decoupling
limit $\phi _{R}^{0}$ corresponds to the $126$~GeV SM-like Higgs boson, i.e $%
h$. The squared masses of the $\phi _{R}^{0}$, $\varphi _{R}^{0}$, $\rho _{R}
$, $\xi _{R}$ scalars are given by:

\begin{eqnarray}
m_{h}^{2} &=&2\lambda _{1}v^{2},\hspace{0.7cm}\hspace{0.7cm}m_{\varphi
_{R}^{0}}^{2}=\mu _{3}^{2}+\frac{1}{2}\left( \lambda _{5}+\lambda
_{8}+\lambda _{11}\right) v^{2}+\alpha _{17}v_{\chi }^{2}+\alpha
_{18}v_{\zeta }^{2},\hspace{0.7cm}\hspace{0.7cm}  \label{mCPeven} \\
m_{\rho _{R}}^{2} &=&\mu _{6}^{2}+\frac{\alpha _{3}}{2}v^{2}+\kappa
_{12}v_{\chi }^{2}+\kappa _{15}v_{\zeta }^{2}+2\kappa _{17}v_{\zeta }v_{\chi
},\hspace{0.7cm}\hspace{0.7cm}m_{\xi _{R}}^{2}=\mu _{7}^{2}+\frac{\alpha _{4}%
}{2}v^{2}+\kappa _{13}v_{\chi }^{2}+\kappa _{16}v_{\zeta }^{2}+2\kappa
_{18}v_{\zeta }v_{\chi }.  \notag
\end{eqnarray}

The scalar fields $H_{1}$ and $H_{2}$ are physical mass eigenstates of the
following squared scalar mass matrix written in the $\left( \widetilde{\chi }%
,\widetilde{\zeta }\right) $ basis:

\begin{equation}
M_{H}^{2}=\left( 
\begin{array}{cc}
8\kappa _{4}v_{\chi }^{2} & 4\kappa _{10}v_{\chi }v_{\zeta } \\ 
4\kappa _{10}v_{\chi }v_{\zeta } & 8\kappa _{5}v_{\zeta }^{2}%
\end{array}%
\right) ,
\end{equation}

This matrix can be diagonalized as follows:

\begin{eqnarray}
R_{H}^{T}M_{H}^{2}R_{H} &=&\left( 
\begin{array}{cc}
4\kappa _{4}v_{\chi }^{2}+4\kappa _{5}v_{\zeta }^{2}+4\sqrt{\left( \kappa
_{4}v_{\chi }^{2}-\kappa _{5}v_{\zeta }^{2}\right) ^{2}+\kappa _{10}v_{\chi
}^{2}v_{\zeta }^{2}} & 0 \\ 
0 & 4\kappa _{4}v_{\chi }^{2}+4\kappa _{5}v_{\zeta }^{2}-4\sqrt{\left(
\kappa _{4}v_{\chi }^{2}-\kappa _{5}v_{\zeta }^{2}\right) ^{2}+\kappa
_{10}v_{\chi }^{2}v_{\zeta }^{2}}%
\end{array}%
\right) ,  \notag \\
R_{H} &=&\left( 
\begin{array}{cc}
\cos \theta _{H} & -\sin \theta _{H} \\ 
\sin \theta _{H} & \cos \theta _{H}%
\end{array}%
\right) \hspace{0.7cm}\hspace{0.7cm}\tan 2\theta _{H}=\frac{\kappa
_{10}v_{\chi }v_{\zeta }}{\kappa _{4}v_{\chi }^{2}-\kappa _{5}v_{\zeta }^{2}}%
.
\end{eqnarray}

Consequently, the physical scalar mass eigenstates states of the matrix $%
M_{H}^{2}$ are given by: 
\begin{equation}
\left( 
\begin{array}{c}
H_{1} \\ 
H_{2}%
\end{array}%
\right) =\left( 
\begin{array}{cc}
\cos \theta _{H} & \sin \theta _{H} \\ 
-\sin \theta _{H} & \cos \theta _{H}%
\end{array}%
\right) \left( 
\begin{array}{c}
\widetilde{\chi } \\ 
\widetilde{\zeta }%
\end{array}%
\right) .
\end{equation}

Their squared masses are: 
\begin{equation}
m_{H_{1/2}}^{2}=4\kappa _{4}v_{\chi }^{2}+4\kappa _{5}v_{\zeta }^{2}\pm 4%
\sqrt{\left( \kappa _{4}v_{\chi }^{2}-\kappa _{5}v_{\zeta }^{2}\right)
^{2}+\kappa _{10}v_{\chi }^{2}v_{\zeta }^{2}}.
\end{equation}

The scalar fields $S_{1}$ and $S_{2}$ are physical mass eigenstates of the
following squared scalar mass matrix written in the $\left( \eta
_{R}^{0},\sigma _{R}\right) $ basis:

\begin{equation}
M_{H}^{2}=\left( 
\begin{array}{cc}
\mu _{2}^{2}+\frac{1}{2}\left( \lambda _{4}+\lambda _{7}+\lambda
_{10}\right) v^{2}+\alpha _{15}v_{\chi }^{2}+\alpha _{16}v_{\zeta }^{2} & 
\frac{1}{\sqrt{2}}Av \\ 
\frac{1}{\sqrt{2}}Av & \mu _{4}^{2}+2\mu _{5}^{2}+\frac{1}{2}\left( \alpha
_{1}+2\alpha _{2}\right) v^{2}+\kappa _{11}v_{\chi }^{2}+\kappa
_{14}v_{\zeta }^{2}%
\end{array}%
\right) ,
\end{equation}

This matrix can be diagonalized as follows:

\begin{eqnarray}
R_{S}^{T}M_{S}^{2}R_{S} &=&\left( 
\begin{array}{cc}
\frac{A_{S}+B_{s}}{2}-\frac{1}{2}\sqrt{\left( A_{S}-B_{S}\right)
^{2}+4C_{S}^{2}} & 0 \\ 
0 & \frac{A_{S}+B_{S}}{2}+\frac{1}{2}\sqrt{\left( A_{S}-B_{S}\right)
^{2}+4C_{S}^{2}}%
\end{array}%
\right) ,  \notag  \label{eq:Theta-S} \\
R_{S} &=&\left( 
\begin{array}{cc}
\cos \theta _{S} & -\sin \theta _{S} \\ 
\sin \theta _{S} & \cos \theta _{S}%
\end{array}%
\right) ,  \notag \\
A_{S} &=&\mu _{2}^{2}+\frac{1}{2}\left( \lambda _{4}+\lambda _{7}+\lambda
_{10}\right) v^{2}+\alpha _{15}v_{\chi }^{2}+\alpha _{16}v_{\zeta }^{2},%
\hspace{0.5cm}\hspace{0.7cm}B_{S}=\mu _{4}^{2}+2\mu _{5}^{2}+\frac{1}{2}%
\left( \alpha _{1}+2\alpha _{2}\right) v^{2}+\kappa _{11}v_{\chi
}^{2}+\kappa _{14}v_{\zeta }^{2},  \notag \\
C_{S} &=&\frac{1}{\sqrt{2}}Av,\hspace{0.7cm}\hspace{0.7cm}\tan 2\theta _{S}=%
\frac{2C_{S}}{A_{S}-B_{S}}.
\end{eqnarray}

Consequently, the physical scalar mass eigenstates states of the matrix $%
M_{S}^{2}$ are given by: 
\begin{equation}
\left( 
\begin{array}{c}
S_{1} \\ 
S_{2}%
\end{array}%
\right) =\left( 
\begin{array}{cc}
\cos \theta _{S} & \sin \theta _{S} \\ 
-\sin \theta _{S} & \cos \theta _{S}%
\end{array}%
\right) \left( 
\begin{array}{c}
\eta _{R}^{0} \\ 
\sigma _{R}%
\end{array}%
\right) .
\end{equation}

Their squared masses are: 
\begin{equation}
m_{S_{1/2}}^{2}=\frac{A_{S}+B_{S}}{2}\pm \frac{1}{2}\sqrt{\left(
A_{S}-B_{S}\right) ^{2}+4C_{S}^{2}}\;.
\end{equation}

Concerning the CP odd scalar sector, we find that it is composed of one
massless pseudoscalar state, i.e, $\phi _{I}^{0}$, which is identified with
the neutral SM Nambu-Goldstone boson $G_{Z}^{0}$ eaten up by the
longitudinal component of the $Z$ gauge boson, as well as four physical
pseudoscalar fields $\varphi _{I}^{0}$, $\rho _{I}$, $\xi _{I}$, $P_{1}$ and 
$P_{2}$. The squared masses of the $\phi _{I}^{0}$, $\varphi _{I}^{0}$, $%
\rho _{I}$ scalars are given by:

\begin{eqnarray}
m_{\phi _{I}^{0}}^{2} &=&0,\hspace{0.7cm}\hspace{0.7cm}m_{\varphi
_{I}^{0}}^{2}=\mu _{3}^{2}+\frac{1}{2}\left( \lambda _{5}+\lambda
_{8}-\lambda _{11}\right) v^{2}+\alpha _{17}v_{\chi }^{2}+\alpha
_{18}v_{\zeta }^{2},  \notag \\
m_{\rho _{I}}^{2} &=&\mu _{6}^{2}+\frac{\alpha _{3}}{2}v^{2}+\kappa
_{12}v_{\chi }^{2}+\kappa _{15}v_{\zeta }^{2}-2\kappa _{17}v_{\zeta }v_{\chi
},\hspace{0.7cm}\hspace{0.7cm}m_{\xi _{I}}^{2}=\mu _{7}^{2}+\frac{\alpha _{4}%
}{2}v^{2}+\kappa _{13}v_{\chi }^{2}+\kappa _{16}v_{\zeta }^{2}-2\kappa
_{18}v_{\zeta }v_{\chi }.  \label{mCPodd}
\end{eqnarray}%
The scalar fields $P_{1}$ and $P_{2}$ are the physical mass eigenstates of
the following squared scalar mass matrix written in the $\left( \eta
_{I}^{0},\sigma _{I}\right) $-basis:

\begin{equation}
M_{P}^{2}=\left( 
\begin{array}{cc}
\mu _{2}^{2}+\frac{1}{2}\left( \lambda _{4}+\lambda _{7}-\lambda
_{10}\right) v^{2}+\alpha _{15}v_{\chi }^{2}+\alpha _{16}v_{\zeta }^{2} & -%
\frac{1}{\sqrt{2}}Av \\ 
-\frac{1}{\sqrt{2}}Av & \mu _{4}^{2}-2\mu _{5}^{2}+\frac{1}{2}\left( \alpha
_{1}-2\alpha _{2}\right) v^{2}+\kappa _{11}v_{\chi }^{2}+\kappa
_{14}v_{\zeta }^{2}%
\end{array}%
\right) ,
\end{equation}%
which can be diagonalized by the transformation: 
%
\begin{eqnarray}
R_{P}^{T}M_{P}^{2}R_{P} &=&\left( 
\begin{array}{cc}
\frac{A_{P}+B_{P}}{2}+\frac{1}{2}\sqrt{\left( A_{P}-B_{P}\right)
^{2}+4C_{P}^{2}} & 0 \\ 
0 & \frac{A_{P}+B_{P}}{2}-\frac{1}{2}\sqrt{\left( A_{P}-B_{P}\right)
^{2}+4C_{P}^{2}}%
\end{array}%
\right) ,  \notag  \label{eq:Theta-P} \\
R_{P} &=&\left( 
\begin{array}{cc}
\cos \theta _{P} & -\sin \theta _{P} \\ 
\sin \theta _{P} & \cos \theta _{P}%
\end{array}%
\right) ,  \notag \\
A_{P} &=&\mu _{2}^{2}+\frac{1}{2}\left( \lambda _{4}+\lambda _{7}-\lambda
_{10}\right) v^{2}+\alpha _{15}v_{\chi }^{2}+\alpha _{16}v_{\zeta }^{2},%
\hspace{0.5cm}\hspace{0.7cm}B_{P}=\mu _{4}^{2}-2\mu _{5}^{2}+\frac{1}{2}%
\left( \alpha _{1}-2\alpha _{2}\right) v^{2}+\kappa _{11}v_{\chi
}^{2}+\kappa _{14}v_{\zeta }^{2},  \notag \\
C_{P} &=&-\frac{1}{\sqrt{2}}Av,\hspace{0.7cm}\hspace{0.7cm}\tan 2\theta _{P}=%
\frac{2C_{P}}{A_{P}-B_{P}}.
\end{eqnarray}%
Consequently, the physical scalar mass eigenstates $P_{1,2}$ 
are given by: 
\begin{equation}
\left( 
\begin{array}{c}
P_{1} \\ 
P_{2}%
\end{array}%
\right) =\left( 
\begin{array}{cc}
\cos \theta _{P} & \sin \theta _{P} \\ 
-\sin \theta _{P} & \cos \theta _{P}%
\end{array}%
\right) \left( 
\begin{array}{c}
\eta _{I}^{0} \\ 
\sigma _{I}%
\end{array}%
\right) .
\end{equation}%
Their squared masses are: 
\begin{equation}
m_{P_{1}}^{2}=\frac{A_{P}+B_{P}}{2}+\frac{1}{2}\sqrt{\left(
A_{P}-B_{P}\right) ^{2}+4C_{P}^{2}},\hspace{0.7cm}\hspace{0.7cm}%
m_{P_{2}}^{2}=\frac{A_{P}+B_{P}}{2}-\frac{1}{2}\sqrt{\left(
A_{P}-B_{P}\right) ^{2}+4C_{P}^{2}}.
\end{equation}

In the charged scalar sector we find two massless Nambu-Goldstone states, $%
\phi^{\pm }$, absorbed by the longitudinal components of $W^{\pm }$ gauge
bosons, as well as four physical charged scalars, $\eta ^{\pm }$, $\varphi
^{\pm }$ with the masses:

\begin{equation}
m_{\eta ^{\pm }}^{2}=\mu _{2}^{2}+\frac{1}{2}\lambda _{4}v^{2}+\alpha
_{15}v_{\chi }^{2}+\alpha _{16}v_{\zeta }^{2},\hspace{0.7cm}\hspace{0.7cm}%
m_{\varphi ^{\pm }}^{2}=\mu _{3}^{2}+\frac{1}{2}\lambda _{5}v^{2}+\alpha
_{17}v_{\chi }^{2}+\alpha _{18}v_{\zeta }^{2}.
\end{equation}%
This completes the list of the scalar sector of our model.

\section{SM Fermion mass hierarchy}

\label{fermionmasshierarchy} The SM fermion mass matrices are generated in
our model according to the diagrams in Fig.~\ref{Loopdiagrams}, with the
Yukawa interactions in (\ref{lyu})-(\ref{lnu}). We write the mass matrices
for the charged fermions in the form 
\begin{eqnarray}
&&M_{U}=\left( 
\begin{array}{ccc}
\left( a_{11}^{\left( u\right) }\right) ^{3}l^{2} & \left( a_{12}^{\left(
u\right) }\right) ^{2}l & a_{13}^{\left( u\right) } \\ 
\left( a_{21}^{\left( u\right) }\right) ^{3}l^{2} & \left( a_{22}^{\left(
u\right) }\right) ^{2}l & a_{23}^{\left( u\right) } \\ 
\left( a_{31}^{\left( u\right) }\right) ^{3}l^{2} & \left( a_{32}^{\left(
u\right) }\right) ^{2}l & a_{33}^{\left( u\right) }%
\end{array}%
\right) \allowbreak \allowbreak \frac{v}{\sqrt{2}},\hspace{0.5cm}%
M_{D}=\left( 
\begin{array}{ccc}
\left( a_{11}^{\left( d\right) }\right) ^{3}l^{2} & \left( a_{12}^{\left(
d\right) }\right) ^{2}l^{2} & \left( a_{13}^{\left( d\right) }\right) ^{2}l
\\ 
\left( a_{21}^{\left( d\right) }\right) ^{3}l^{2} & \left( a_{22}^{\left(
d\right) }\right) ^{2}l^{2} & \left( a_{23}^{\left( d\right) }\right) ^{2}l
\\ 
\left( a_{31}^{\left( d\right) }\right) ^{3}l^{2} & \left( a_{32}^{\left(
d\right) }\right) ^{2}l^{2} & \left( a_{33}^{\left( d\right) }\right) ^{2}l%
\end{array}%
\right) \allowbreak \allowbreak \frac{v}{\sqrt{2}}  \label{eq:Qmass-Matrices}
\\
&&M_{l}=\left( 
\begin{array}{ccc}
\left( a_{11}^{\left( l\right) }\right) ^{3}l^{2} & \left( a_{12}^{\left(
d\right) }\right) ^{2}l & \left( a_{13}^{\left( l\right) }\right) ^{2}l \\ 
\left( a_{21}^{\left( l\right) }\right) ^{3}l^{2} & \left( a_{22}^{\left(
d\right) }\right) ^{2}l & \left( a_{23}^{\left( l\right) }\right) ^{2}l \\ 
\left( a_{31}^{\left( l\right) }\right) ^{3}l^{2} & \left( a_{32}^{\left(
d\right) }\right) ^{2}l & \left( a_{33}^{\left( l\right) }\right) ^{2}l%
\end{array}%
\right) \allowbreak \allowbreak \frac{v}{\sqrt{2}}
\label{eq:Lepton-Mass-Matrix}
\end{eqnarray}%
%
%
%
Here we have taken into account the loop level at which the columns of these
matrices are generated, in particular $l\approx (1/4\pi )^{2}$ is the loop
suppression factor.

The powers of this loop factor in (\ref{eq:Qmass-Matrices}), (\ref%
{eq:Lepton-Mass-Matrix}), explicitly display the following picture that we
have in our model: The third column in $M_U$ is generated at the tree-level,
engendering mass to the top quark. The second and first columns of $M_U$
arise at the one and two-loop levels, respectively, and are associated with
the charm and up quark masses. The light down and strange quark masses are
also generated at two-loop level. On the other hand, the third column of $%
M_D $ arise at one-loop level.

As for the SM charged lepton mass matrix $M_{l}$, its first column,
responsible for the electron mass, appears at the two-loop level, whereas
its second and third columns, providing masses to the muon and the tau
lepton, respectively, are generated at one loop.

As we pointed out before, 
the objective of the model is to generate the observed hierarchy of the
fermion mass spectrum in terms of loop suppression. Therefore, it is crucial
that the quark masses and mixings predicted by the model are reproduced with
parameters $a_{ij}^{\left( u\right) }$, $a_{ij}^{\left(d\right) }\sim 
\mathcal{O}(1)$ ($i,j=1,2,3$), 
Let us check this essential point in detail: We use the experimental values
of the quark masses \cite{Xing:2019vks}, the CKM parameters \cite%
{Tanabashi:2018oca} and the charged lepton masses \cite{Tanabashi:2018oca}: 
\begin{eqnarray}  \label{eq:Qsector-observables}
&& m_{u}(MeV)=1.24\pm 0.22, \hspace{3mm} m_{d}(MeV)=2.69\pm 0.19, \hspace{3mm%
} m_{s}(MeV)=53.5\pm 4.6,  \notag  \label{eq:Qsector-observables} \\
&&m_{c}(GeV)=0.63\pm 0.02,\hspace{3mm} m_{t}(GeV)=172.9\pm 0.4,\hspace{3mm}
m_{b}(GeV)=2.86\pm 0.03,\hspace{3mm}  \notag \\
&&\sin\theta_{12}=0.2245\pm 0.00044,\hspace{3mm} \sin \theta_{23}=0.0421\pm
0.00076,\hspace{3mm} \sin \theta_{13}=0.00365\pm 0.00012,  \notag \\
&&J=\left(3.18\pm 0.15\right)\times 10^{-5}\,, \\
&& m_{e}(MeV) = 0.4883266\pm 0.0000017, \ \ \ \ m_{\mu}(MeV) =102.87267\pm
0.00021,\ \ \ \ m_{\tau}(MeV) =1747.43\pm 0.12,  \notag
\end{eqnarray}
where, $J$ is the Jarlskog parameter.

By solving the eigenvalue problem for the mass matrices (\ref%
{eq:Qmass-Matrices}), (\ref{eq:Lepton-Mass-Matrix}) we find a solution for
the parameters that reproduces the values in Eq.~(\ref%
{eq:Qsector-observables}). It is given by 
\begin{eqnarray}
a_{ij}^{\left( u\right) }&=&\left( 
\begin{array}{ccc}
-0.688435 & 0.23427 & 0.574417 \\ 
-0.433888 & 0.975784 & 0.575768 \\ 
0.460125 & 0.299329 & 0.572606 \\ 
&  & 
\end{array}
\right), \\
a_{ij}^{\left( d\right) }&=&\left( 
\begin{array}{ccc}
0.496199\, -0.856786 i & 0.553843\, -0.956252 i & 0.988976\, +0.00132749 i
\\ 
0.0000811073\, -0.9244 i & 0.000107414\, -1.13112 i & 0.924773\,
+0.0000767587 i \\ 
0.00207731\, +0.985775 i & 0.00249437\, +1.15427 i & 0.987132\, -0.00207885 i
\\ 
&  & 
\end{array}
\right)\,, \\
a_{ij}^{\left( l\right) }&=&\left( 
\begin{array}{ccc}
-0.598992-0.00493263 i & 0.00393775\, -0.916528 i & 0.77355\, +0.00318633 i
\\ 
0.000405292\, +0.675959 i & 0.000325396\, +0.880957 i & 0.676159\,
-0.000407163 i \\ 
0.00295785\, +0.801275 i & 0.0036143\, +0.898469 i & 0.801517\, -0.0029589 i
\\ 
&  & 
\end{array}
\right)\,.  \notag
\end{eqnarray}
As we can see, all the entries (the absolute values) of the above matrices
are of order unity with rather mild deviations. This demonstrates that the
proposed model is able to explain the existing pattern of the observed quark
spectrum. via the sequential loop suppression mechanism. 

Finally, the small masses of the active neutrinos are generated at the
three-loop level, as follows from the last diagram of Figure~\ref%
{Loopdiagrams}. Thus, for the neutrino mass matrix we can write %
\begin{equation}
M_{\nu} = l^3 y^6 \lambda\frac{v^2}{M},  \label{estimate}
\end{equation}
with $M$ denoting a common mass scale of the virtual scalars and fermions
running in the internal lines of the neutrino loop diagram in Figure~\ref%
{Loopdiagrams}, $y$ is a matrix of the neutrino Yukawa couplings and $%
\lambda $ is the quartic scalar coupling. Using $M\sim\mathcal{O}%
\left(13\right)$ TeV, 
$y\sim 0.3 \mathbbm{1}$, and $\lambda\sim 0.1$ in Eq. (\ref{estimate}) we
find $m_{\nu}\sim\mathcal{O}\left(0.1\right)$ eV, thus showing that the
model naturally explains the smallness of the light active neutrino masses
with respect to the EWSB scale. 
Furthermore, from this estimate of the light neutrino masses it is to expect
that exotic scalars and fermions beyond the SM should have masses of~$%
\mathcal{O}\left(13\right)$~TeV.

\section{Charged lepton-flavor violation constraints}

\label{LFV} 
\begin{figure}[!h]
\vspace{-1cm} \includegraphics[width=1.3\textwidth]{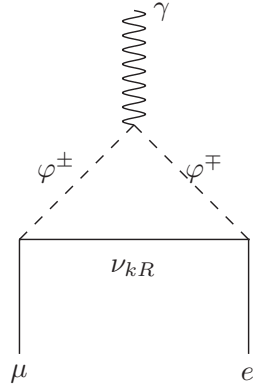}%
\vspace{-22cm}
\caption{Feynman diagram corresponding to the dominant contribution to
the $\protect\mu\to e\protect\gamma$ decay.}
\label{Diagmutoegamma}
\end{figure}

\begin{figure}[tbp]
\includegraphics[width=0.8 \textwidth]{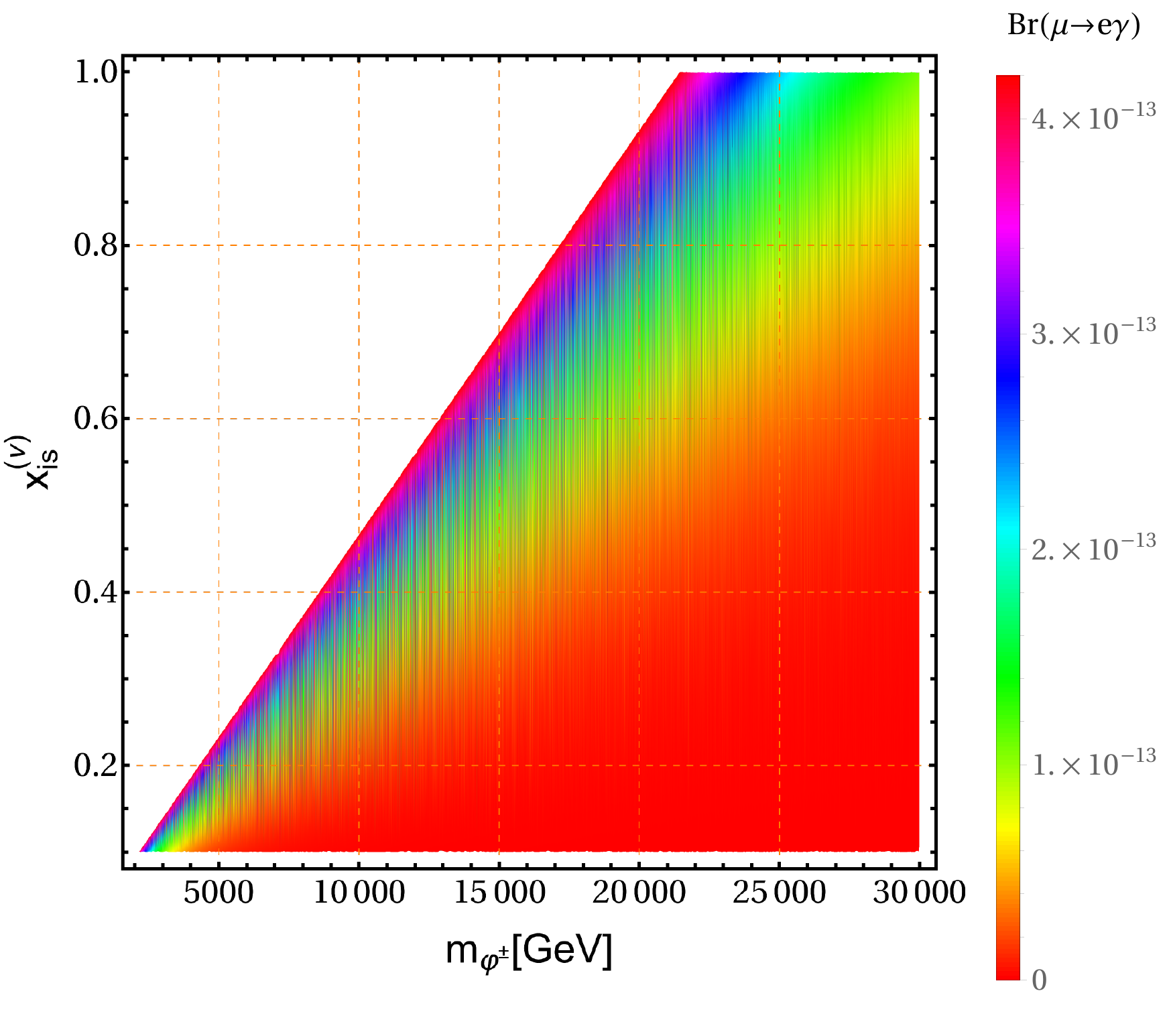}
\caption{Allowed parameter space in the $m_{\protect\varphi ^{\pm
}}-x_{js}^{\left( \protect\nu \right) }$ plane consistent with the charged
lepton flavor-violating constraints.}
\label{LFVplot}
\end{figure}

\begin{figure}[tbp]
\includegraphics[width=0.8 \textwidth]{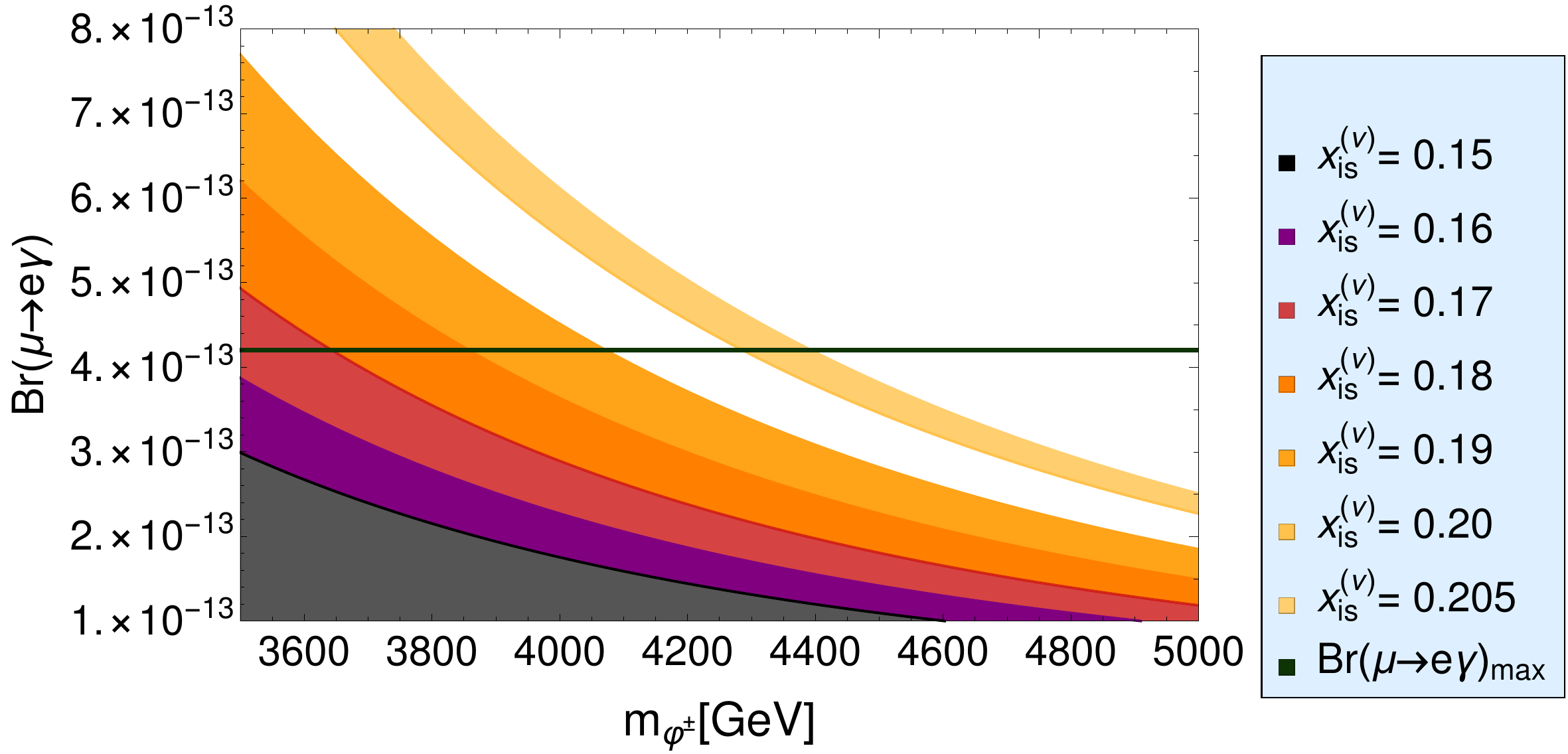}
\caption{Branching ratio for the $\protect\mu\to e\protect\gamma$ decay as
function of charged scalar masses $m_{\protect\varphi ^{\pm }}$ for
different values of the $x_{js}^{\left( \protect\nu \right) }$ couplings.
The horizontal line corresponds to the experimental upper bound of $%
4.2\times 10^{-13}$ \protect\cite{Lindner:2016bgg} for the branching ratio
of the $\protect\mu \rightarrow e\protect\gamma $ decay. Here we have set $%
m_N=50$ MeV}
\label{mutoegamma}
\end{figure}

In this section we will derive constraints 
from the non-observation of the charged Lepton Flavor Violating (LFV)
process $\mu \rightarrow e\gamma$. The dominant contribution to the decay $%
l_{i}\rightarrow l_{j}\gamma$ occurs in our model at
one-loop level and, according to the diagram in Figure~\ref{Diagmutoegamma},
is mediated by a virtual electrically charged scalar $\varphi ^{+}$,
originating from the $SU(2)_{L}$ inert doublet $\varphi $, and by the
right-handed Majorana neutrinos \mbox{$\nu _{sR}$ ($s=1,2$)}. There is also
a contribution arising from the charged exotic leptons $E_{2}$ and the
electrically neutral scalars $S_{k}, P_{k}$. However this contribution is
sub-leading since it only appears at the two loop level, as shown in
Appendix~\ref{sec:ESP-MUEgamma}. 
Therefore, we can safely neglect it. 
Then we find for the branching ratio corresponding to the diagram
in Fig.~\ref{Diagmutoegamma} the following expression \cite{Ma:2001mr,Toma:2013zsa,Vicente:2014wga,Lindner:2016bgg} 
\begin{eqnarray}
Br\left( l_{i}\rightarrow l_{j}\gamma \right) &=&\frac{3\left( 4\pi \right)
^{3}\alpha _{em}}{4G_{F}^{2}}\left\vert \sum_{s=1}^{2}\frac{x_{is}^{\left(
\nu \right) }x_{js}^{\left( \nu \right) }}{2\left( 4\pi \right)
^{2}m_{\varphi ^{\pm }}^{2}}F\left( \frac{m_{\nu _{sR}}^{2}}{m_{\varphi
^{\pm }}^{2}}\right) \right\vert ^{2}Br\left( l_{i}\rightarrow l_{j}\nu _{i}%
\overline{\nu _{j}}\right) ,  \notag \\
F\left( x\right) &=&\frac{1-6x+3x^{2}+2x^{3}-6x^{2}\ln x}{6\left( 1-x\right)
^{4}}\;.  \label{BRclfv}
\end{eqnarray}
Here $x_{is}^{\left(\nu \right)}=\sum_{k=1}^{3}y_{ks}^{\left(\nu
\right)}\left( V_{lL}^{\dagger}\right) _{ik}$ and $m_{\varphi ^{\pm }}$ are
the masses of the charged scalar components of the $SU(2)_{L}$ inert doublet 
$\varphi $, whereas $m_{\nu _{sR}}$ ($s=1,2$) correspond to the masses of
the right-handed Majorana neutrinos $\nu _{sR}$. 
To simplify our analysis we choose a benchmark scenario where the
right-handed Majorana neutrinos $\nu _{sR}$ are all degenerate with respect
to a common mass $m_{N}$. In our numerical analysis we vary these masses in
the following ranges 
\mbox{$1$ TeV$\leqslant m_{\varphi ^{\pm}}\leqslant 30$
TeV} and \mbox{$10$ MeV$\lesssim m_{N}\leqslant 100$ MeV}. We also vary the
dimensionless couplings in the window $0.1\leqslant x_{is}^{\left(
\nu\right) }\leqslant 1$ ($i=1,2,3$ and $s=1,2$). Let us note that we
scanned only over the MeV scale masses for the right-handed Majorana
neutrinos $\nu_{sR}$, since these masses are generated at the two-loop
level, as seen from the two-loop sub-diagram of the third Feynman diagram of
Fig.~\ref{Loopdiagrams}. This is the same loop level at which masses of
light and strange quarks, lying in the MeV region, are generated. The
results of our analysis are displayed in Figures~\ref{LFVplot} and~\ref%
{mutoegamma}. In Figure~\ref{LFVplot} we plot the allowed parameter space in
the $m_{\varphi ^{\pm }}-x_{js}^{\left( \nu \right) }$ plane consistent with
the existing $\mu\rightarrow e\gamma$ experimental constraints. This plot is
obtained by randomly generating the parameters $m_{N}$, $m_{\varphi ^{\pm }}$%
, $x_{is}^{\left( \nu \right) }$ and $x_{js}^{\left( \nu \right) }$ in a
range of values where the $\mu \rightarrow e\gamma $ branching ratio is
below its upper experimental limit of $4.2\times 10^{-13}$ \cite%
{Lindner:2016bgg}. As can be seen from Figure \ref{LFVplot}, this condition
is satisfied for the charged scalar masses $m_{\varphi ^{\pm }}$ larger than
about $3.5$ TeV. We also find that in the same region of parameter space,
our model predicts branching ratios for the $\tau\rightarrow \mu \gamma $
and $\tau \rightarrow e\gamma $ decays up to $10^{-10}$, which is below
their corresponding upper experimental bounds of $4.4\times 10^{-9}$ and $%
3.3\times 10^{-9}$, respectively. Consequently, the model is compatible with
the current charged lepton-flavor-violating decay constraints. The branching
ratio for the $\mu\to e\gamma$ decay as a function of the charged scalar
mass $m_{\varphi ^{\pm }}$ is shown in Fig.~\ref{mutoegamma} for different
values of the $x_{js}^{\left( \nu \right) }$ couplings. This Figure shows
that the branching ratio for the $\mu\to e\gamma$ decay decreases as the
charged scalar masses $m_{\varphi ^{\pm }}$ acquire larger values. The
horizontal line corresponds to the experimental upper bound of $4.2\times
10^{-13}$ \cite{Lindner:2016bgg} for the branching ratio of the $\mu
\rightarrow e\gamma $ decay. Here we set $m_N=50$~MeV. We have checked that
the branching ratio for the $\mu\to e\gamma$ decay has a very low
sensitivity to the mass $m_{N}$ of the right-handed Majorana neutrinos $\nu
_{sR}$ ($s=1,2$). 

Given that future experiments, such as Mu2e and COMET~\cite{Bernstein:2013hba},
are expected to measure or at least constrain lepton-flavor conversion in
nuclei with much better precision than the radiative lepton LFV decays, we
proceed to derive constraints imposed on the model parameter space by $\mu-e$
conversion in nuclei. The $\mu^{-}-e^{-}$ conversion ratio is defined~\cite%
{Lindner:2016bgg} as: 
\begin{equation}  \label{eq:Conversion-Rate}
{\text{CR}}\left(\mu-e\right)=\frac{\Gamma\left(\mu^{-}+{\text{Nucleus}}%
\left(A,Z\right)\rightarrow e^{-}+{\text{Nucleus}}\left(A,Z\right)\right)}{%
\Gamma\left(\mu^{-}+{\text{Nucleus}}\left(A,Z\right)\rightarrow\nu_{\mu}+{%
\text{Nucleus}}\left(A,Z-1\right)\right)}
\end{equation}
Using an Effective Lagrangian approach for describing LFV processes, as done
in~\cite{Kuno:1999jp}, and considering the low momentum limit, where the
off-shell contributions from photon exchange are negligible with respect to
the contributions arising from real photon emission, the dipole operators
shown in Ref.~\cite{Kuno:1999jp} dominate the conversion rate, thus,
yielding the following relations \cite{Kuno:1999jp,Lindner:2016bgg}: 
\begin{equation}  \label{eq:CR-BR}
{\text{CR}}\left(\mu {\text{Ti}}\rightarrow e{\text{Ti}}\right)\simeq\frac{1%
}{200}{\text{Br}}\left(\mu \rightarrow e\gamma\right)\hspace{1cm}{\text{CR}}%
\left(\mu {\text{Al}}\rightarrow e{\text{Al}}\right)\simeq\frac{1}{350}{%
\text{Br}}\left(\mu \rightarrow e\gamma\right)
\end{equation}
Notice that the aforementioned relations are valid for the case of photon
dominance in the $\mu^{-}-e^{-}$ conversion, which applies to our model due
to the absence of tree-level flavor changing neutral scalar interactions.
Therefore, experimental upper bounds on the conversion rates (\ref%
{eq:Conversion-Rate}) will translate in our model to upper limits on Br$%
(\mu\rightarrow e\gamma)$.

The sensitivity of the CERN Neutrino Factory, which will use a Titanium
target \cite{Aysto:2001zs}, is expected at the level of~$\sim 10^{-18}$. The
expected sensitivities of the next generation experiments such as Mu2e and
COMET \cite{Bernstein:2013hba}, with an Aluminum target, 
are expected to be about~$\sim 10^{-17}$. Thus, according to Eqs.~(\ref%
{eq:CR-BR}), the future limits will result in about three order of magnitude
improvement in Br$(\mu\rightarrow e\gamma)$.

\begin{figure}[tbp]
\includegraphics[width=0.8\textwidth]{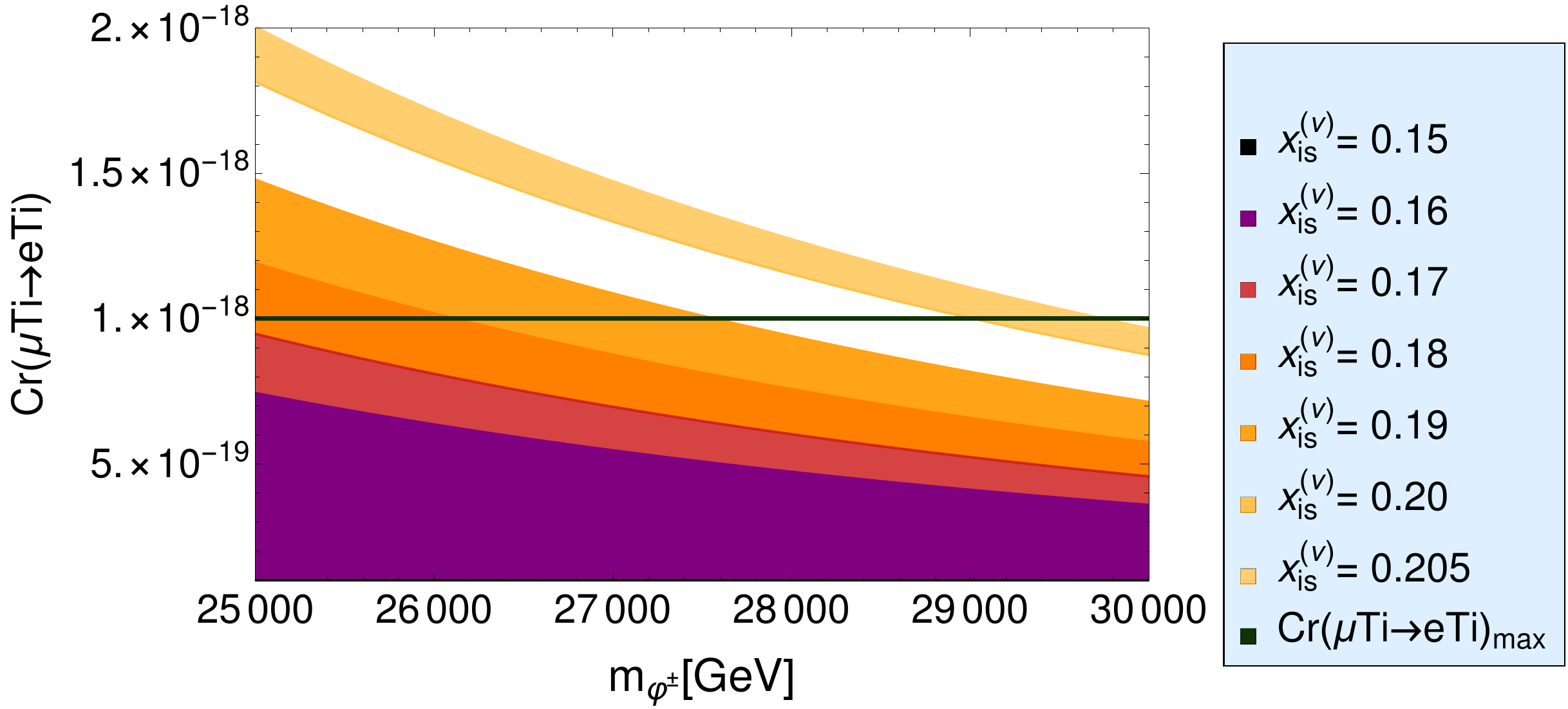}\newline
\includegraphics[width=0.8\textwidth]{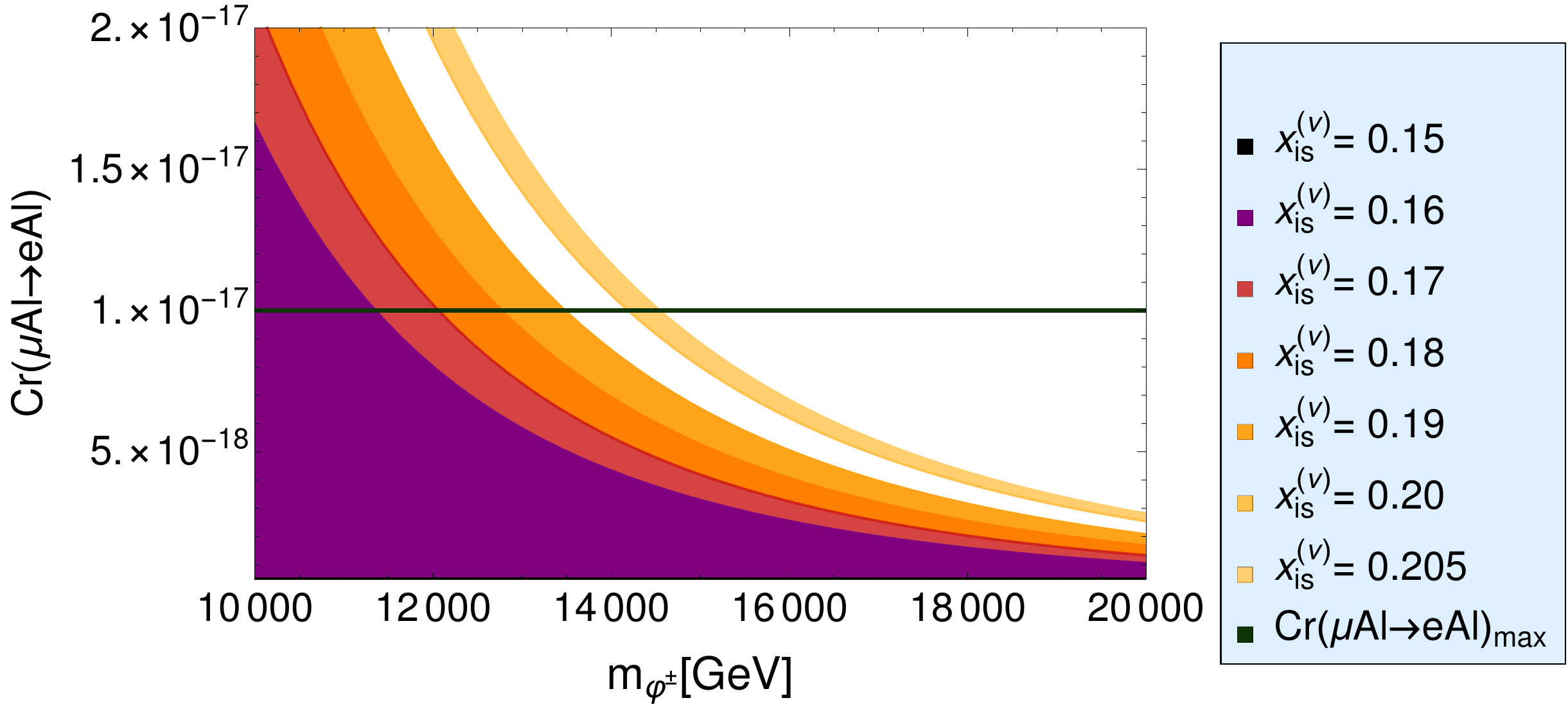}
\caption{CR$\left(\protect\mu {\text{Ti}}\rightarrow e{\text{Ti}}\right)$
(top plot) and CR$\left(\protect\mu {\text{Al}}\rightarrow e{\text{Al}}%
\right)$ (bottom plot) as function of the charged scalar masses $m_{\protect%
\varphi ^{\pm }}$, for different values of the $x_{js}^{\left( \protect\nu %
\right) }$ couplings ($j=1,2,3$, $n=1,2$). The black horizontal line in each
plot corresponds to the expected sensitivities of the next generation of
experiments that will use titanium~\protect\cite{Aysto:2001zs} and aluminum~%
\protect\cite{Bernstein:2013hba} as targets, respectively. Here we have set $%
m_N=50$~MeV.}
\label{CR}
\end{figure}

In Figure \ref{CR} we show the CR$\left(\mu {\text{Ti}}\rightarrow e{\text{Ti%
}}\right)$ (top plot) and CR$\left(\mu {\text{Al}}\rightarrow e{\text{Al}}%
\right)$ (bottom plot), as function of the charged scalar mass $m_{\varphi
^{\pm }}$ for different values of the dimensionless couplings $%
x_{js}^{\left( \nu \right) }$ ($j=1,2,3$, $n=1,2$). The black horizontal
lines correspond to the expected sensitivities $\sim 10^{-18}$ (top plot) of
the CERN Neutrino Factory~\cite{Aysto:2001zs} and $\sim 10^{-17}$ (bottom
plot) of the next generation of experiments such as Mu2e and COMET~\cite%
{Bernstein:2013hba}. In these plots we have set $m_N=50$~MeV. The plots show
that the next generation experiments, where titanium and aluminium will be
used as targets, can rule out the part of the model parameter space where
the charged scalar masses are lower than about $10$~TeV for $x_{js}^{\left(
\nu \right) }\simeq\mathcal{O}(0.1)$. 

\section{Muon and electron anomalous magnetic moment.}

\label{gminus2mu} 
\begin{figure}[!h]
\vspace{-1cm} \includegraphics[width=1.5\textwidth]{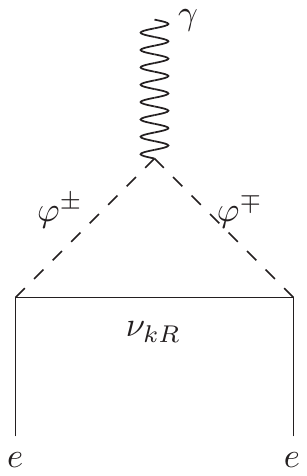}%
\vspace{-26cm}
\caption{Leading Feynman-loop diagram contributing to the electron anomalous
magnetic moment. Here $k=1,2$.}
\label{G2electron}
\end{figure}

The results of the experimental measurements of the
anomalous magnetic dipole moments of electron and muon $a_{e,\mu }=(g_{e,\mu
}-2)/2$ show significant deviation from their SM values 
\begin{eqnarray}
\Delta a_{\mu } &=&a_{\mu }^{\mathrm{exp}}-a_{\mu }^{\mathrm{SM}}=\left(
2.51\pm 0.59\right) \times 10^{-9}\hspace{17mm}\mbox{%
\cite{Hagiwara:2011af,Davier:2017zfy,Nomura:2018lsx,Nomura:2018vfz,Blum:2018mom,Keshavarzi:2018mgv,Aoyama:2020ynm,Abi:2021gix}}
\label{eq:a-mu} \\
\Delta a_{e} &=&a_{e}^{\mathrm{exp}}-a_{e}^{\mathrm{SM}}= (-0.88\pm
0.36)\times 10^{-12}\hspace{3mm}\mbox{\cite{Parker:2018vye}}, \hspace{3mm}
(4.8\pm 3.0)\times 10^{-13}\hspace{3mm}\mbox{\cite{Morel:2020dww}}
\label{eq:a-e}
\end{eqnarray}%
%
%
%
%
%
%
%
%
%
%
%
%
%
%
%
%
%
%
%
%
%
Here the value of $a^{\mathrm{exp}}_{\mu}$ is a combined result of the BNL
E821 experiment \cite{Bennett:2006fi} and the recently announced FNAL Muon
g-2 measurement \cite{Abi:2021gix} , showing the 4.2$\sigma$ tension between
the SM and experiment. 
%
The last positive value for $\Delta a_{e}$ corresponds to the recently
published new measurement of the fine-structure constant with an accuracy of
81 parts per trillion~\cite{Morel:2020dww}. In this section we analyze
predictions of our model for these observables. The leading contributions to 
$\Delta a_{e,\mu }$ arising in the model are shown in Figs.~\ref%
{Diaggminus2muon}, \ref{G2electron}.

\begin{figure}[!h]
\vspace{-1cm} \includegraphics[width=1.3\textwidth]{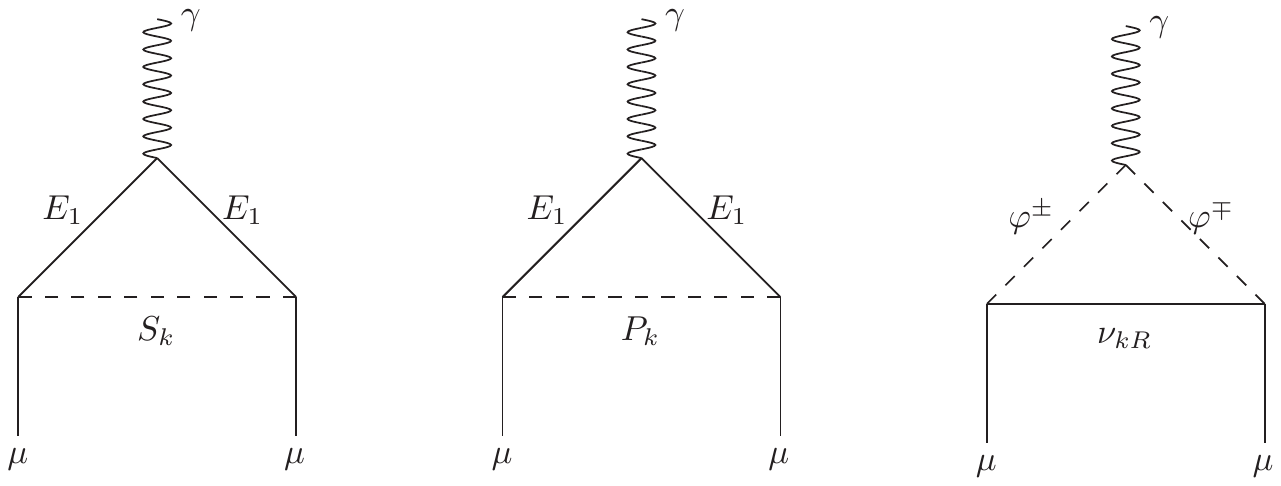}%
\vspace{-22cm}
\caption{Feynman-loop diagrams contributing to the muon anomalous magnetic
moment. Here $k=1,2$.}
\label{Diaggminus2muon}
\end{figure}
\begin{figure}[!h]
\includegraphics[width=0.9\textwidth]{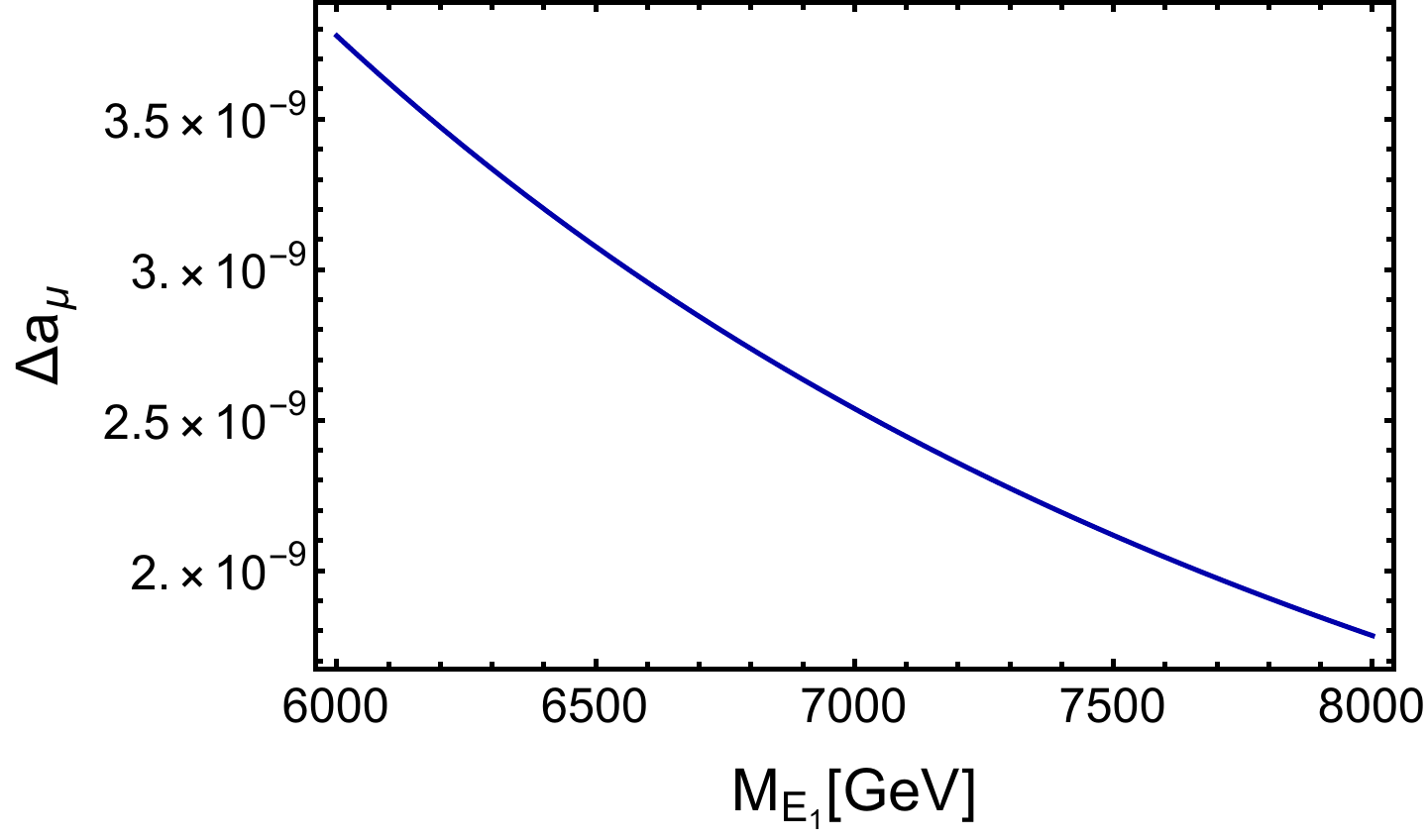}
\caption{Muon anomalous magnetic moment as a function of the charged exotic
lepton mass $M_{E_2}$.}
\label{gminus2muonvsmE}
\end{figure}

For simplicity we set $\theta _{S}=\theta_{P}=\theta$ and $y_{22}^{\left(
l\right) }=x_{22}^{\left( l\right) }=y_{21}^{\left( \nu \right)
}=y_{22}^{\left( \nu \right) }=y$ (for the definitions see Eqs. (\ref{lyl}),
(\ref{lnu}), (\ref{eq:Theta-S}) and (\ref{eq:Theta-P})). Furthermore, we
work on a simplified benchmark scenario with a diagonal SM charged lepton
mass matrix, where the charged exotic leptons $\tilde{E}$, $\widetilde{E}^{\prime }$; $E_1$ and $E_2$,
only contribute to the electron, muon and tau masses, respectively. Then,
the contribution to the muon anomalous magnetic moment takes the form 
\begin{eqnarray}
\Delta a_{\mu } &=&\frac{y^{2}m_{\mu }^{2}}{8\pi ^{2}}\left[ I_{S}\left(
m_{E_{1}},m_{S_{1}}\right) -I_{S}\left( m_{E_{1}},m_{S_{2}}\right)
+I_{P}\left( m_{E_{1}},m_{P_{1}}\right) -I_{P}\left(
m_{E_{1}},m_{P_{2}}\right) \right] \sin \theta \cos \theta  \notag \\
&&-\frac{y^{2}_2m_{\mu }^{2}}{16\pi ^{2}m_{\varphi ^{\pm }}^{2}}%
\sum_{s=1}^{2}F\left( \frac{m_{\nu _{sR}}^{2}}{m_{\varphi ^{\pm }}^{2}}%
\right),  \label{deltaamu}
\end{eqnarray}
where the loop integral $F\left(x\right)$ is defined in Eq. (\ref{BRclfv})
and was previously computed in Ref.~\cite{Ma:2001mr}, whereas $I_{S\left(
P\right) }\left( m_{E},m\right) $ has the form \cite%
{Diaz:2002uk,Jegerlehner:2009ry,Kelso:2014qka,Lindner:2016bgg,Kowalska:2017iqv}

\begin{equation}
I_{S\left( P\right) }\left( m_{E},m\right) =\int_{0}^{1}\frac{x^{2}\left(
1-x\pm \frac{m_{E}}{m_{\mu }}\right) }{m_{\mu }^{2}x^{2}+\left(
m_{E}^{2}-m_{\mu }^{2}\right) x+m^{2}\left( 1-x\right) }dx.  \label{I}
\end{equation}%
In our numerical analysis we consider a benchmark scenario with $\theta =%
\frac{\pi }{4}$, $m_{\nu _{1R}}=m_{\nu _{2R}}=50$~MeV, $M_{P_1}=M_{S_1}=0.5$%
~TeV, $M_{P_2}=0.6$~TeV, $M_{S_2}=1$ TeV and $m_{\varphi ^{+}}=4$ TeV. The
mass of the charged exotic lepton $E_1$ has been varied in the ranges $6$ TeV%
$\leqslant M_{E_1}\leqslant$ $8$ TeV. Note that these masses for the
right-handed Majorana neutrinos $\nu _{sR}$ ($s=1,2$) and for the
electrically charged scalar $\varphi ^{+}$ are consistent with the
constraints arising from the charged lepton flavor processes $\mu
\rightarrow e\gamma $, $\tau \rightarrow \mu \gamma $ and $\tau \rightarrow
e\gamma $, as shown in the previous section. Considering that the muon
anomalous magnetic moment is constrained to be in the range shown in (\ref%
{eq:a-mu}), we plot in Fig.~\ref{gminus2muonvsmE} the muon anomalous
magnetic moment as a function of the charged exotic lepton mass $M_{E_1}$.
Figure \ref{gminus2muonvsmE} shows that the muon anomalous magnetic moment
decreases when the charged exotic lepton mass is increased.

The anomalous magnetic moment of the electron $\Delta a_{e}$ can be computed
in an analogous way as $\Delta a_{\mu }$. The difference is that the neutral
(pseudo-)scalars and exotic charged leptons contribution to the $\Delta
a_{e} $ appears at two-loop level, as shown in Appendix~\ref%
{sec:ESP-MUEgamma}, and is therefore sub-leading. Thus, $\Delta a_{e}$ is
dominated by the effective vertex diagram in Fig~\ref{G2electron}, involving
the electrically charged scalar $\varphi ^{+}$, which couples to the
right-handed Majorana neutrinos $\nu _{sR}$ ($s=1,2$). From this diagram we
find in an approximate form~\cite{Ma:2001mr} 
\begin{eqnarray}
\Delta a_{e} &\approx&-\frac{y^{2}_1m_{e}^{2}}{16\pi ^{2}m_{\varphi ^{\pm
}}^{2}}\sum_{s=1}^{2}F\left( \frac{m_{\nu _{sR}}^{2}}{m_{\varphi ^{\pm }}^{2}%
}\right).  \label{deltaae}
\end{eqnarray}
Consequently, our model predicts negative values for this observable in
accordance with~\cite{Parker:2018vye}. However, in order to reproduce either 
\cite{Parker:2018vye} or \cite{Morel:2020dww}, the experimental values shown
in~(\ref{eq:a-e}), we need that the mass of the electrically charged scalar $%
\varphi ^{\pm }$ lies in the interval 
\mbox{$100$ GeV $\lesssim m_{\varphi
^{\pm }}\lesssim$ $150$ GeV}. These values are incompatible with the $%
\mu\rightarrow e\gamma$ constraints analyzed in Section~\ref{LFV}. The
latter require $m_{\varphi ^{\pm }}\geq 3.5$ TeV, which will yield in this
case a bit too small value for the electron anomalous magnetic moment, which
nonetheless are consistent with the above mentioned $2\sigma$ experimentally
allowed range.

\section{Dark Matter relic density}

\label{darkmatter} In this section we provide a discussion of our model in
view of Dark Matter (DM). We do not intend to provide a sophisticated
analysis of the DM constraints, which is beyond the scope of the present
paper. Note that due to the preserved $Z_{2}^{\left( 2\right) }$ discrete
symmetry and to the residual $Z_{2}$ symmetry (arising from the spontaneous
breaking of the $Z_{4}$ subgroup), our model has several stable scalar DM
candidates.
 As follows from
the scalar assignments to the $Z_{2}^{\left( 2\right)
}\times Z_{4}$ symmetry, given by Eq. (\ref{assignmentsscalars}), we can
assign this role to any of the following scalar particles: $\varphi
_{R}^{0}$, $\rho _{R}$, $\xi _{R}$, $S_{1}$, $S_{2}$, $\varphi _{I}^{0}$, $%
\rho _{I}$, $\xi _{I}$, $P_{1}$ or $P_{2}$. Furthermore, our model has a
fermionic dark matter candidate, which can be the lightest among the two
right-handed Majorana neutrinos $\nu _{sR}$ ($s=1,2$), since in our model
they are the only right-handed Majorana neutrinos whose masses appear at
two-loop level.

Based on Eqs. (\ref{mCPeven}) and (\ref{mCPodd}) of Section \ref%
{scalarpotential}, we take $\rho _{I}$ as our scalar Dark Matter candidate.
To guarantee the stability of $\rho _{I}$, we assume that this field is
lighter than the charged exotic fermions, and in this way its decay modes
into exotic and SM charged fermions are kinematically forbidden.

The relic density of the Dark Matter in the present Universe is estimated as
follows (c.f. Ref.~\cite{Tanabashi:2018oca,Edsjo:1997bg}) 
\begin{equation}
\Omega h^{2}=\frac{0.1\;\text{pb}}{\left\langle \sigma v\right\rangle },\,%
\hspace{1cm}\left\langle \sigma v\right\rangle =\frac{A}{n_{eq}^{2}}\,,
\end{equation}%
where $\left\langle \sigma v\right\rangle $ is the thermally averaged
annihilation cross section, $A$ is the total annihilation rate per unit
volume at temperature $T$ and $n_{eq}$ is the equilibrium value of the
particle density, which are given in \cite{Edsjo:1997bg} 
\begin{eqnarray}
A &=&\frac{T}{32\pi ^{4}}\dint\limits_{4m_{\varphi }^{2}}^{\infty
}\dsum\limits_{p=W,Z,t,b,h}g_{p}^{2}\frac{s\sqrt{s-4m_{\rho _{I}}^{2}}}{2}%
v_{rel}\sigma \left( \rho _{I}\rho _{I}\rightarrow p\overline{p}\right)
K_{1}\left( \frac{\sqrt{s}}{T}\right) ds,  \notag \\
n_{eq} &=&\frac{T}{2\pi ^{2}}\dsum\limits_{p=W,Z,t,b,h}g_{p}m_{\rho
_{I}}^{2}K_{2}\left( \frac{m_{\rho _{I}}}{T}\right) ,
\end{eqnarray}%
with $K_{1}$ and $K_{2}$ being the modified Bessel functions of the second
kind of order 1 and 2, respectively \cite{Edsjo:1997bg}. For the relic
density calculation, we take $T=m_{\rho _{I}}/20$ as in Ref. \cite%
{Edsjo:1997bg}, which corresponds to a typical freeze-out temperature.

The scalar DM candidate $\rho _{I}$ annihilates mainly into $WW$, $ZZ$, $t%
\overline{t}$, $b\overline{b}$ and $hh$, via a Higgs portal scalar
interaction $\left( \phi ^{\dagger }\phi \right) \rho _{I}\rho _{I}$, where $%
\phi $ is the SM Higgs doublet. The corresponding annihilation cross
sections are given by: \cite{Bhattacharya:2016ysw}: 
\begin{eqnarray}
v_{rel}\sigma \left( \rho _{I}\rho _{I}\rightarrow WW\right) &=&\frac{\alpha
_{3}}{32\pi }\frac{s\left( 1+\frac{12m_{W}^{4}}{s^{2}}-\frac{4m_{W}^{2}}{s}%
\right) }{\left( s-m_{h}^{2}\right) ^{2}+m_{h}^{2}\Gamma _{h}^{2}}\sqrt{1-%
\frac{4m_{W}^{2}}{s}},  \notag \\
v_{rel}\sigma \left( \rho _{I}\rho _{I}\rightarrow ZZ\right) &=&\frac{\alpha
_{3}}{64\pi }\frac{s\left( 1+\frac{12m_{Z}^{4}}{s^{2}}-\frac{4m_{Z}^{2}}{s}%
\right) }{\left( s-m_{h}^{2}\right) ^{2}+m_{h}^{2}\Gamma _{h}^{2}}\sqrt{1-%
\frac{4m_{Z}^{2}}{s}},  \notag \\
v_{rel}\sigma \left( \rho _{I}\rho _{I}\rightarrow q\overline{q}\right) &=&%
\frac{N_{c}\alpha _{3}^{2}m_{q}^{2}}{16\pi }\frac{\sqrt{\left( 1-\frac{%
4m_{f}^{2}}{s}\right) ^{3}}}{\left( s-m_{h}^{2}\right) ^{2}+m_{h}^{2}\Gamma
_{h}^{2}},  \notag \\
v_{rel}\sigma \left( \rho _{I}\rho _{I}\rightarrow hh\right) &=&\frac{\alpha
_{3}^{2}}{64\pi s}\left( 1+\frac{3m_{h}^{2}}{s-m_{h}^{2}}-\frac{2\alpha
_{3}v^{2}}{s-2m_{h}^{2}}\right) ^{2}\sqrt{1-\frac{4m_{h}^{2}}{s}},
\end{eqnarray}%
where $\sqrt{s}$ is the centre-of-mass energy, $N_{c}=3$ is the color
factor, $m_{h}=125.7$ GeV and $\Gamma _{h}=4.1$ MeV are the SM Higgs boson $%
h $ mass and its total decay width, respectively; $\alpha _{3}$ is the
quartic scalar coupling corresponding to the interaction $\alpha _{3}\left(
\phi ^{\dagger }\phi \right) \left( \rho ^{\dagger }\rho \right) $.

\begin{figure}[t]
\center
\vspace{0.8cm}\includegraphics[width=0.7\textwidth]{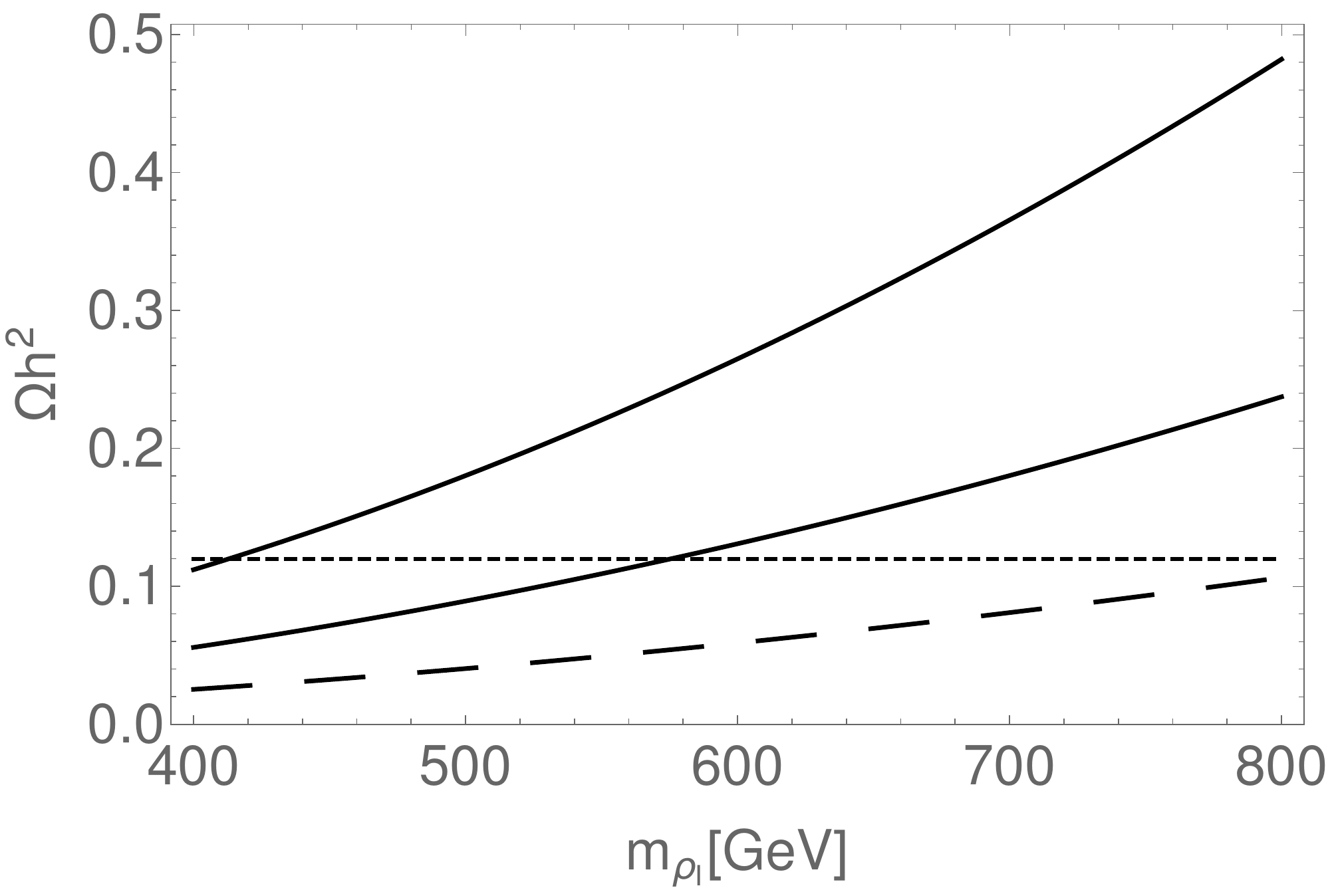}
\caption{Relic density $\Omega h^{2}$, as a function of the mass $m_{\protect%
\rho _{I}}$ of the $\protect\rho _{I}$ scalar field, for several values of
the quartic scalar coupling $\protect\alpha_3$. The curves from top to
bottom correspond to $\protect\alpha_3=1,1.2,1.5$, respectively. The
horizontal line shows the observed value $\Omega h^{2}=0.1198$ \protect\cite%
{Ade:2015xua} for the relic density.}
\label{DM}
\end{figure}

\begin{figure}[t]
\center
\vspace{0.8cm} \includegraphics[width=0.7\textwidth]{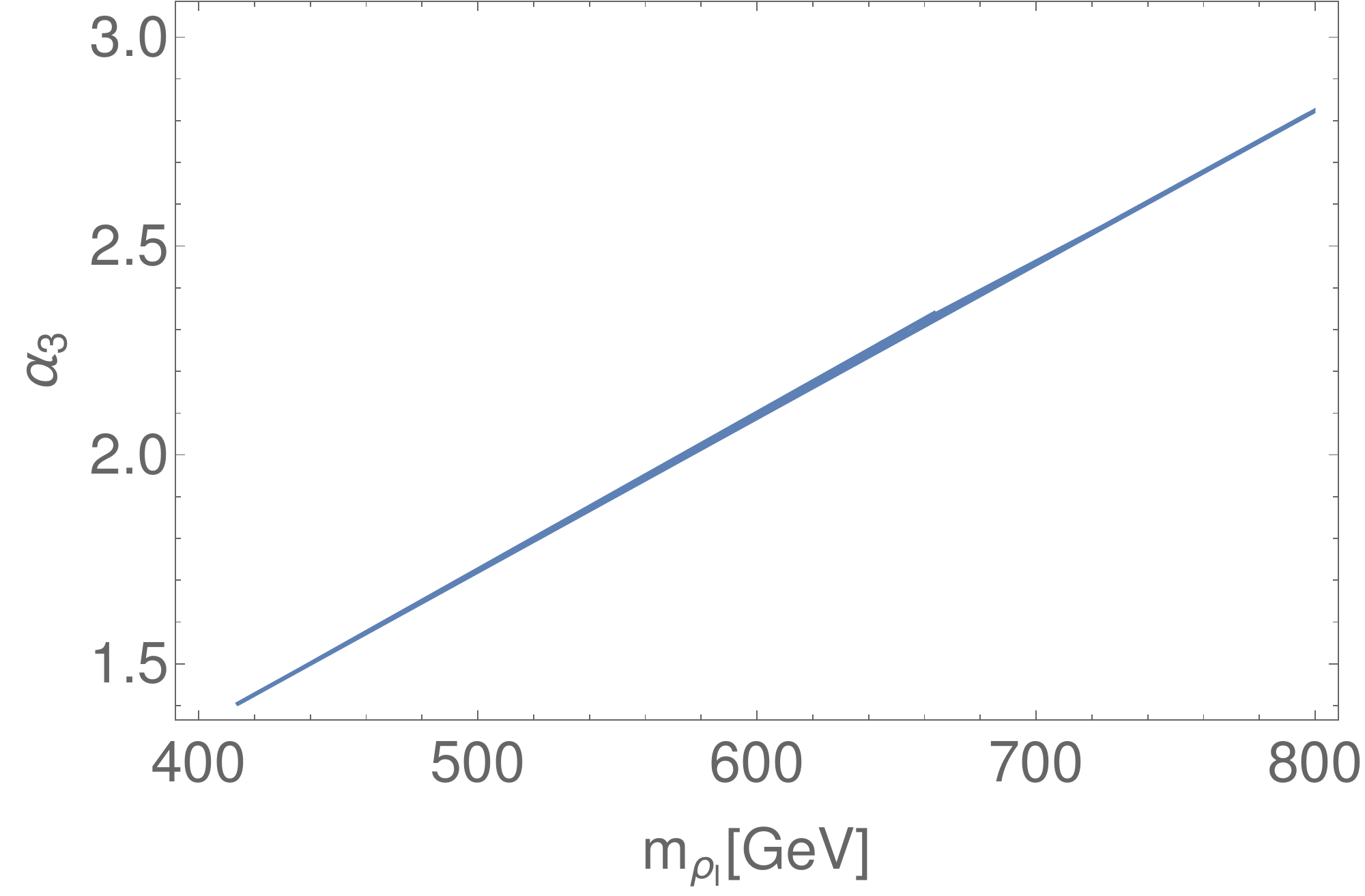}
\caption{Correlation between the quartic scalar coupling $\protect\alpha_3$
and the mass $m_{\protect\rho _{I}}$ of the scalar Dark Matter candidate $%
\protect\rho _{I}$, consistent with the experimental value $\Omega
h^{2}=0.1198$ for the Relic density.}
\label{CorrelationDM}
\end{figure}

Fig.~\ref{DM} displays the relic density $\Omega h^{2}$ as a function of the
mass $m_{\rho _{I}}$ of the scalar field $\rho _{I}$, for several values of
the quartic scalar coupling $\alpha _{3}$. The curves from top to bottom
correspond to $\alpha _{3}$ =1, 1.2 and 1.5, respectively. The horizontal
line corresponds to the experimental value $\Omega h^{2}=0.1198$ of the
relic density. Figure \ref{DM} shows that the relic density is an increasing
function of the mass $m_{\varphi }$ and a decreasing function of the quartic
scalar coupling $\alpha _{3}$. Consequently, an increase in the mass $%
m_{\rho _{I}}$ of the scalar field $\rho _{I}$ will require a larger quartic
scalar coupling $\alpha _{3}$, in order to account for the measured value of
the Dark Matter relic density, as indicated in Fig.~\ref{CorrelationDM}.

It is worth mentioning that the Dark Matter relic density constraint yields
a linear correlation between the quartic scalar coupling $\alpha_3$ and the
mass $m_{\rho _{I}}$ of the scalar Dark Matter candidate $\rho _{I}$, as
shown in Fig.~\ref{CorrelationDM}. We have numerically checked that in order
to reproduce the observed value, $\Omega h^{2}=0.1198\pm 0.0026$ \cite%
{Ade:2015xua}, of the relic density, the mass $m_{\rho _{I}}$ of the scalar
field $\rho _{I}$\ has to be in the range $400$ GeV$\leqslant m_{\rho
_{I}}\leqslant$ $800$~GeV, for a quartic scalar coupling $\alpha _{3}$ in
the range $1\leqslant \alpha _{3}\leqslant 1.5$. 

In what concerns prospects for the direct DM detection, the scalar DM
candidate would scatter off a nuclear target in a detector via Higgs boson
exchange in the $t$-channel, giving rise to a constraint on the coupling of
the $\left( \phi ^{\dagger }\phi \right) \rho _{I}\rho _{I}$ interaction.

\section{Conclusions}

\label{conclusions} We have constructed an extension of the 3HDM based on
the $Z_{2}^{\left( 1\right) }\times Z_{2}^{\left( 2\right) }\times Z_{4}$
symmetry, where the SM particle content is enlarged by two inert $SU_{2L}$
scalar doublets, three inert and two active electrically neutral gauge
singlet scalars, charged vector like fermions and Majorana neutrinos. These
fields are introduced in order to generate the SM fermion mass hierarchy
from a sequential loop suppression mechanism: tree-level top quark mass;
1-loop bottom, charm, tau and muon masses; 2-loop masses for the light up,
down and strange quarks as well as for the electron; and 3-loop masses for
the light active neutrinos. In our model, the $Z_{2}^{\left(
2\right) }$ symmetry is preserved, whereas the $Z_{2}^{\left( 1\right) }$
symmetry is completely broken and the $Z_{4}$ symmetry is broken down to a
conserved $Z_{2}$ symmetry, thus allowing the stability of the Dark Matter
as well as a successful implementation of the aforementioned sequential loop
suppression mechanism, without the inclusion of soft symmetry breaking terms.
For studying the electroweak symmetry breaking in our model we applied the
bilinear formalism of the 3HDM.

We demonstrated that our model successfully accommodates the current fermion
mass spectrum and fermionic mixing parameters, the electron and muon
anomalous magnetic moments, as well as the constraints arising from charged
lepton flavor violating processes. 

We have also shown that in our model the branching ratios of the decays 
$\mu\to e\gamma$, $\tau\rightarrow \mu
\gamma $ and $\tau \rightarrow e\gamma $ can reach values of the order 
of $10^{-13}$, which is within the
reach of the future experimental sensitivity, thus making our model testable
by the forthcoming experiments.

Finally, we have examined the scalar DM
particle candidate of the model and have shown that the prediction is
compatible with the observed DM relic density abundance for scalar masses in
the range $400$ GeV$\leqslant m_{\rho_{I}}\leqslant $ $800$ GeV.

\section*{Acknowledgments}

A.E.C.H, S.K., M.M, and I.S. are supported by ANID-Chile FONDECYT 1210378,
ANID-Chile FONDECYT 1190845, ANID-Chile FONDECYT 1200641, ANID-Chile
FONDECYT 1180232, ANID-Chile FONDECYT 3150472, ANID PIA/APOYO AFB180002 and
Milenio-ANID-ICN2019\_044.

\appendix

\section{Exotic Leptons and Neutral scalar contribution to Leptonic LFV
decays}

\label{sec:ESP-MUEgamma} Let us show that the contribution to $%
l_{i}\rightarrow l_{j}\gamma$ decay of the charged exotic leptons $E_{2}$
and the electrically neutral scalars $S_{k}, P_{k}$ vanishe at one loop.
Their one-loop contribution is given by the first two diagrams in Fig.~\ref%
{Diaggminus2muon}, with one $\mu$ replaced by $e$.

In the mass eigenstate basis $\widetilde{l}_i$ the corresponding
contribution to the branching fraction is given by: 
\begin{eqnarray}
{\text{Br}}\left( \widetilde{l}_{a}\rightarrow \widetilde{l}_{b}\gamma
\right)^{\text{1-loop}} _{\text{scalar}} &\simeq &\kappa
\sum_{j=1}^{3}\sum_{k=1}^{3}\left( V_{lL}^{\dagger }\right) _{aj}\left[
\sum_{s=1}^{2}y_{js}^{\left( l\right) }x_{sk}^{\left( l\right) }\left(
\delta _{k2}+\delta _{k3}\right) \right] \left( V_{lR}\right) _{kb}\left(
1-\delta _{ab}\right) F_{1loop}  \notag \\
&=&\frac{\kappa }{G_{1loop}\frac{v}{\sqrt{2}}}\sum_{j=1}^{3}\sum_{k=1}^{3}%
\left( V_{lL}^{\dagger }\right) _{aj}M_{jk}^{\left( l\right) }\left(
V_{lR}\right) _{kb}\left( 1-\delta _{ab}\right) F_{1loop}  \notag \\
&=&\frac{\kappa }{G_{1loop}\frac{v}{\sqrt{2}}}\sum_{j=1}^{3}\sum_{k=1}^{3}%
\left( m_{\mu }\delta _{a2}\delta _{b2}+m_{\tau }\delta _{a3}\delta
_{b3}\right) \left( 1-\delta _{ab}\right) F_{1loop}=0,
\end{eqnarray}
what was to be shown. Here we have taken into account that the SM charged
lepton mass matrix has the form:

\begin{eqnarray}
M_{jk}^{\left( l\right) }=\left[ \sum_{s=1}^{2}y_{js}^{\left( l\right)
}x_{sk}^{\left( l\right) }\left( \delta _{k2}+\delta _{k3}\right)
G_{1loop}+y_{j}^{\left( l\right) }x_{1}^{\left( l\right) }\delta
_{k1}G_{2loop}\right] \frac{v}{\sqrt{2}}\simeq\sum_{s=1}^{2}y_{js}^{\left(
l\right) }x_{sk}^{\left( l\right) }\left( \delta _{k2}+\delta
_{k3}\right)G_{1loop}\frac{v}{\sqrt{2}}  \label{Mlcompact}
\end{eqnarray}
and satisfies 
\begin{equation}
V_{lL}^{\dagger }M^{\left( l\right) }V_{lR}=\left( M^{\left( l\right)
}\right)_{diag}
\end{equation}
where $j,k=1,2,3$, with $G_{1loop}$ and $G_{2loop}$ being the corresponding
one and two loop functions, respectively.

The SM fermionic fields in the mass ($\widetilde{f}_{\left( L,R\right) }$)
and interaction ($f_{\left( L,R\right) }$) eigenstate bases are related as 
\begin{equation}
f_{\left( L,R\right) }=V_{f\left( L,R\right) }\widetilde{f}_{\left(
L,R\right)}\;.  \label{fandftilde}
\end{equation}
\vspace{2cm}

\centerline{\bf{REFERENCES}}\vspace{-0.4cm} 
\bibliographystyle{utphys}
\bibliography{biblio19thApril2021}

\end{document}